\documentclass[a4paper,11pt]{article}
\usepackage{jheppub}
\usepackage{verbatim}
\usepackage{mdframed}
\definecolor{lightgraybg}{gray}{0.945}
\usepackage{booktabs} 
\usepackage{xcolor}
\usepackage{listings}
\usepackage{booktabs}
\usepackage{arydshln}
\usepackage{mdframed}
\usepackage{xcolor}
\usepackage{listings}
\usepackage{hyperref}
\definecolor{lightgraybg}{RGB}{248,248,248}

\lstdefinestyle{terminal}{
    basicstyle=\ttfamily\small,
    columns=fullflexible,
    keepspaces=true,
    breaklines=true,
    showstringspaces=false,
    frame=none,
    xleftmargin=0pt,
    aboveskip=0pt,
    belowskip=0pt
}
\title{Language-Guided Hypotheses Generation for Sparse SMEFT Analyses}

\author[a]{Ahmed Hammad}
\author[b]{ and  Veronica Sanz}

\affiliation[a]{Center of AI and Natural Sciences, KIAS, Seoul 02455, Korea}
\affiliation[b]{IFIC, Universitat de València-CSIC, Valencia, Spain}

\emailAdd{hammad@kias.re.kr}
\emailAdd{veronica.sanz@uv.es}

\abstract{
Global fits of the Standard Model Effective Field Theory are challenged by the large number of operators, while any given database constrains only a small subset. Selecting relevant operator hypotheses therefore requires theoretical insight into operator correlations and the sensitivity of observables. We present \texttt{llm4smeft}, an open source framework that addresses this problem by combining a language model, fine-tuned on the SMEFT literature, with retrieval augmented generation based on quantitative summaries of \texttt{SMEFiT} package global fits. Given a set of observables, the framework proposes candidate relevant operators together with their corresponding Fisher information, while retrieval ensures that model outputs are grounded in existing fit results whenever available. The framework runs in an interactive mode in which accepted hypotheses are stored in a growing knowledge base. We publicly release the \texttt{llm4smeft}  package together with the fine-tuned language model\footnote{\href{https://github.com/vsanz/llm4smeft}{Framework: GitHub}

~~\href{https://huggingface.co/ahammad115566/qwen-smeft}{Finetuned LLM: qwen-smeft}
}, in which  the entire framework runs locally, requiring neither internet access nor paid cloud services.
}

\begin{document}
\maketitle
\section{Introduction}
The Standard Model Effective Field Theory (SMEFT) has become a central framework for interpreting precision measurements and searches for physics beyond the Standard Model in a model-independent manner~\cite{Brivio:2017vri,Isidori:2023pyp}. By encoding the low-energy effects of heavy new physics through higher-dimensional operators, it provides a common language for combining information from a wide range of observables, including electroweak precision measurements, Higgs production and decay rates, diboson processes, top-quark observables and flavour constraints~\cite{Ellis:2018gqa,Ellis:2020unq,Ethier:2021bye,Celada:2024mcf,Mantani:2025qji}. The increasing precision of current experiments, together with the anticipated sensitivity of future facilities, has transformed SMEFT analyses into high-dimensional inference problems involving large operator bases and non-trivial correlations among datasets.\\ 
While modern fitting frameworks can constrain dozens of Wilson coefficients simultaneously, the interpretation of fit results remains a challenging task. In practice, phenomenologists often begin by examining which datasets exhibit the largest tensions with the Standard Model (SM) and then attempt to identify a small subset of operators capable of explaining the observed pattern. For example, improvements concentrated in electroweak precision observables may point toward operators associated with oblique corrections~\cite{Bagnaschi:2022whn}, while correlated deviations in Higgs and diboson measurements may suggest modifications of gauge--Higgs interactions. Similarly, tensions appearing primarily in top-quark observables may motivate the consideration of top-sector operators. Such reasoning is routinely employed to formulate simplified hypotheses, guide theoretical investigations, and prioritize directions for more detailed numerical studies. 

The rapid growth of the SMEFT parameter space makes this process increasingly difficult. Even when attention is restricted to a limited subset of operators, the number of possible combinations grows combinatorially, and distinct operator directions can often produce similar phenomenological signatures. Consequently, identifying sparse and physically motivated hypotheses frequently requires substantial expert knowledge and repeated exploration of a large theory space before dedicated numerical fits can be carried out. Data-driven model-selection strategies, including evolutionary searches over operator subsets, provide one route through this combinatorial problem~\cite{Hirsch:2025qxx}, while machine-learning methods have also been developed to retain multivariate information in global SMEFT analyses~\cite{GomezAmbrosio:2022mpm}.\\
Recent advances in large language models (LLMs) suggest a complementary strategy for addressing such complex tasks. In high-energy physics, their applications have developed along three main directions. The first focuses on adapting foundation models to the domain and evaluating their capabilities for scientific reasoning and knowledge retrieval~\cite{Zhang:2024xiwu,Hellert:2024physbert,Barman:2025lpm,Chung:2025tpbench,Barman:2025benchmark,Rafique:2026dune,Richmond:2025lzg,Heneka:2025fpe,Lu:2025izv,Cai:2026ths}. The second, and currently most active, direction targets analysis automation, where single- and multi-agent systems plan analyses, generate and validate code, delegate numerical calculations to established software and reproduce published studies~\cite{Plehn:2026gxv,Diefenbacher:2025zzn,Esmail:2026jpb,Birk:2026zpd,Gendreau-Distler:2025fsj,Desai:2026nmx,Bakshi:2025fgx,Menzo:2025cim,Qiu:2026iby,Agrawal:2026lvg,Faroughy:2026dkj,Costa:2026oew,Wang:2026jjn,Menzo:2026qrl,Niarchos:2026fbt,Hammad:2026ged,Lucente:2026kgh,Guo:2026ifi,Moreno:2026jfc}. A third and more recent direction shifts the emphasis from generating code to generating scientific hypotheses, using LLMs to propose physically motivated explanations for experimental observations~\cite{Diefenbacher:2026azr,Alexander:2026lpw,Saad:2026pan}. Our work belongs to this latter category. We investigate whether an LLM can serve as a theory-informed prior over which SMEFT operators are worth fitting, conditioned on the outcome of existing global analyses while leaving all numerical inference to established fitting tools. 

In this work, we investigate whether large language models can assist SMEFT analyses by acting as theory guided proposal engines. Rather than performing parameter estimation or statistical inference, our framework generates candidate operator hypotheses from compact summaries of experimental information. The objective is to infer a small set of operators that could plausibly account for an observed pattern of deviations, thereby reducing the dimensionality of the hypothesis space explored by subsequent fitting procedures. \\
More specifically, we examine whether an LLM can learn meaningful associations between characteristic patterns of experimental behaviour and the SMEFT operators that commonly generate them. For instance, given a summary indicating that electroweak precision observables and Higgs measurements exhibit significant improvements relative to the SM, can the model identify the corresponding gauge--Higgs operator sector as a plausible explanation? More generally, can it recover or infer sparse operator structures without  explicit fitting procedures?

To address these questions, we develop the \texttt{llm4smeft} framework, which combines language model reasoning with precomputed global fits to propose sparse SMEFT operator subsets for a given set of collider observables. The analysis begins by specifying a set of observables exhibiting tension with the SM. From these inputs, the language model constructs an initial case by estimating the significance of each deviation and identifying a candidate set of SMEFT operators that could account for the observed anomalies. The proposed operators are then validated against the implemented operator basis to ensure that only physically meaningful hypotheses are considered.
The retrieved case is then used to generate multiple sparse operator hypotheses, each consisting of a small subset of operators accompanied by a brief explanation of the underlying physics motivation. Restricting the number of operators per hypothesis favors simpler and more interpretable solutions. An option in which the user has to provide the maximum number of the generated hypotheses and the sparsity cap  at each generated hypothesis. In addition, the framework supports an interactive workflow in which the user may accept the proposed hypotheses or provide additional feedback to refine the language model answer. Based on the feedback from the user, the framework may refine its answer multiple times until the user accept the generated hypotheses.  Once the user accepts the final result, the accepted hypotheses are stored as a new record in the same format as the fit-derived summaries, restricted to the observable groups that were requested, so that they are available for retrieval in subsequent sessions.

The remainder of this paper is organized as follows. In Section~2, we formulate the problem. Section~3 presents the proposed framework, while Section~4 validates and assesses its performance. Our conclusions are presented in Section~5. Installation instructions and a quick start guide for the package are provided in the appendix.

\section{Problem formulation}
\label{sec:problem}

SMEFT provides a model independent framework to describe possible effects of physics beyond the SM. It extends the SM Lagrangian by adding the complete set of gauge invariant operators constructed from Standard Model fields and organized according to their canonical dimension. Retaining only the dimension six operators, which provide the leading corrections for most observables at current experimental precision, the Lagrangian is written as
\begin{equation}
\mathcal{L}_{\rm SMEFT} = \mathcal{L}_{\rm SM} + \sum_i \frac{c_i}{\Lambda^2} \mathcal{O}_i + \mathcal{O}(\Lambda^{-4})\,,
\label{eq:lagrangian}
\end{equation}
where $\Lambda$ denotes the scale of new physics and the Wilson coefficients $c_i$ parameterize its low-energy effects. The baryon-number-conserving dimension-six Warsaw basis was established in Ref.~\cite{Grzadkowski:2010es}; once flavour indices are resolved, it contains 2499 independent operators~\cite{Alonso:2013hga}. Under the flavour assumptions commonly adopted in global analyses, however, the number of relevant coefficients is reduced to only a few tens~\cite{Brivio:2017vri,Fuentes-Martin:2020zaz}.
Experimental data constrain the predictions of the theory rather than the Lagrangian itself. For a given observable, the SMEFT prediction depends polynomially on the Wilson coefficients,
\begin{equation}
\mathcal{T}(c) = \mathcal{T}^{\rm SM}
\left(1 + \sum_{i=1}^{n_{\rm op}} \frac{c_i}{\Lambda^2}\, a_i
        + \sum_{i,j=1}^{n_{\rm op}} \frac{c_i c_j}{\Lambda^4}\, b_{ij}\right),
\label{eq:prediction}
\end{equation}
where the coefficients $a_i$ arise from the interference between a single dimension-six SMEFT amplitude and the SM amplitude, while $b_{ij}$ describe the squared dimension-six contribution and the interference among dimension-six amplitudes. The sums run over the $n_{\rm op}$ Wilson coefficients retained in the analysis. At the same order in $\Lambda^{-1}$, a complete EFT prediction would in general also contain the interference of dimension-eight amplitudes with the SM; these terms are not included here. These coefficients
are computed once using Monte Carlo simulations and stored for each observable and kinematic bin, so that evaluating the likelihood during the fit requires no additional event generation. The Wilson coefficients are determined by comparing these predictions with the experimental measurements. Collecting the $n_{\rm dat}$ measurements in a data vector $D$ and the corresponding predictions in $\mathcal{T}(c)$, the measurements are described by a multivariate Gaussian likelihood,
\begin{equation}
-2\log\mathcal{L}(c) =
\bigl[\mathcal{T}(c) - D\bigr]^{\top} C^{-1} \bigl[\mathcal{T}(c) - D\bigr]
+ \text{const.},
\qquad
C = C^{\rm exp} + C^{\rm th},
\label{eq:likelihood}
\end{equation}
where $C$ is the $n_{\rm dat} \times n_{\rm dat}$ covariance matrix of the measurements, given by the sum of the experimental and theoretical contributions. The theory covariance accounts for missing higher order corrections and provides an important contribution for many observables.
 
At linear order in the SMEFT expansion, Eq.~\eqref{eq:prediction} becomes linear in the Wilson coefficients and can be written as 
\begin{equation}
\mathcal{T}(c) = \mathcal{T}^{\rm SM} + K c\,,
\qquad
K_{ i} = \mathcal{T}^{\rm SM} \frac{a_i}{\Lambda^2}\,,
\label{eq:linear}
\end{equation}
where $K$ is an $n_{\rm dat} \times n_{\rm op}$ matrix whose rows correspond to data points and whose columns correspond to Wilson coefficients.  In this approximation Eq.~\eqref{eq:likelihood} is quadratic in $c$ and can be minimised analytically. Defining the Fisher information matrix as
\begin{equation}
F = K^{\top} C^{-1} K\,,
\label{eq:fisher-def}
\end{equation}
the maximum likelihood estimator of the Wilson coefficients and its covariance
are
\begin{equation}
\hat{c} = F^{-1} K^{\top} C^{-1}
\left( D - \mathcal{T}^{\rm SM} \right),
\qquad
\mathrm{Cov}\!\left(\hat{c}\right) = F^{-1}.
\label{eq:analytic}
\end{equation}
Once the quadratic SMEFT contributions proportional to $b_{ij}$ are
included, the dependence of the theory predictions on the Wilson coefficients
becomes nonlinear. Although Eq.~\eqref{eq:likelihood} retains a Gaussian form in
the residuals, it is no longer Gaussian as a function of the Wilson coefficients,
and no closed-form solution exists. In this case the Wilson coefficients must be
determined using numerical inference methods.

These calculations are implemented in \texttt{SMEFiT}~\cite{Giani:2023gfq}. For each measurement, \texttt{SMEFiT} stores the SM prediction together with the coefficients $a_i$ and $b_{ij}$ of Eq.~\eqref{eq:prediction} in theory tables organized by kinematic bin and perturbative order. A fit is defined through a runcard specifying the datasets, perturbative orders, Wilson coefficients and prior ranges. \\
\texttt{SMEFiT} provides two inference strategies. Linear fits are solved analytically using Eq.~\eqref{eq:analytic}, making them essentially instantaneous. Quadratic fits require nested sampling to explore the non-Gaussian posterior and are therefore considerably more computationally demanding. After the fit, \texttt{SMEFiT} reports the global and per-dataset $\chi^2$, the posterior intervals and correlations of the Wilson coefficients, the decomposition of the Fisher information by dataset, and its principal components. These post fit quantities summarize the information contained in the data and form the basis of the present work.

The analytic solution of Eq.~\eqref{eq:analytic} assumes that the Fisher information matrix is invertible. In practice, the Fisher matrix often exhibits nearly flat directions, reflecting combinations of Wilson coefficients that produce indistinguishable effects in the available observables. Even when it is formally invertible, its eigenvalue spectrum is typically highly hierarchical, indicating that only a small number of parameter combinations are well constrained by the data, while many others remain weakly constrained.

This hierarchy limits the interpretability of fully global SMEFT fits. Marginalizing over poorly constrained directions broadens the bounds on the remaining coefficients and increases their dependence on the assumed priors. For this reason, many SMEFT studies instead consider sparse subsets of operators motivated by specific ultraviolet scenarios or patterns of experimental deviations. Connecting such subsets to ultraviolet dynamics is itself an inverse-matching problem: complete tree-level dictionaries, explicit one-loop matching studies and model-specific analyses show that ultraviolet completions generally populate correlated patterns of coefficients~\cite{deBlas:2017xtg,Brivio:2021yjb,Cepedello:2022fvq}.

\subsection{The combinatorial obstruction}
\label{sec:problem:combinatorics}

Restricting the fit to sparse subsets of operators reduces the number of free parameters, but it introduces a different challenge: identifying which subsets should be tested. If $N$ Wilson  coefficients are available and each hypothesis contains at most $s$ operators, the total number of possible subsets is
\begin{equation}
N_{\rm hyp}(N,s) = \sum_{k=1}^{s} \binom{N}{k}.
\label{eq:combinatorics}
\end{equation}
For the global fit considered in this work, with $N=52$, there are $23\,478$ possible hypotheses already for $s=3$, increasing to $2.9\times10^{5}$ when $s=4$. Exploring all of them is computationally
prohibitive. In particular, each quadratic fit requires an independent nested sampling run, making a large fitting exhaustive.

In practice,  only a small fraction of these subsets are physically interesting. SMEFT experts  can often discard many candidates without performing a fit. Some operators have little sensitivity to the available data, while others are constrained only through specific linear combinations. In addition, theoretical considerations may favor operators that are generated together in well-motivated ultraviolet scenarios or that are relevant to the observed pattern of deviations. Systematic constructions of ultraviolet completions illustrate these correlations for loop-induced four-fermion operators, dark-matter models and neutral triple-gauge interactions~\cite{Cepedello:2022fvq,Cepedello:2023yao,Cepedello:2024fnt}. As a result, selecting a small set of promising hypotheses relies heavily on expert knowledge accumulated
across the SMEFT literature rather than on the outcome of a single global fit.
\subsection{Proposed solution}
\label{sec:problem:approach}

The goal of this work is to automate the selection of sparse SMEFT hypotheses while leaving their evaluation entirely to the statistical fit. Rather than searching exhaustively through all possible operator combinations, we employ a language model specialized
to the SMEFT literature to propose a small number of physically motivated hypotheses for a given set of observables. \\
The framework separates hypothesis generation from statistical inference. The language model is responsible only for proposing candidate operator subsets, while the diagnostics accompanying each proposal are read from pre-generated \texttt{SMEFiT} summaries.
These summaries include the goodness of fit of each dataset, the Fisher information, the dominant constrained directions, and the list of unidentifiable operators. The candidate operators are then ranked according to their Fisher information and restricted to the subset that is most relevant for the requested observables. 

The resulting framework has several desirable properties. It relies on quantitative information extracted from SMEFT fits rather than on the model memory alone, ensuring that identical inputs always produce the same physical context. Retrieval of stored fit summaries is performed before any generation, allowing existing results to be reused whenever possible. Operators that are unconstrained by the selected observables are excluded from the candidate pool, while the Fisher information is used to prioritize the most informative directions in parameter space. Finally, each proposal is reported together with the fit-derived diagnostics of its constituent operators, so that the language model assists the exploration of the parameter space without replacing the statistical inference on which those diagnostics rest.

\section{Methodology}
\label{sec:methodology}

\begin{figure}[!ht]

    \includegraphics[width=0.95\linewidth]{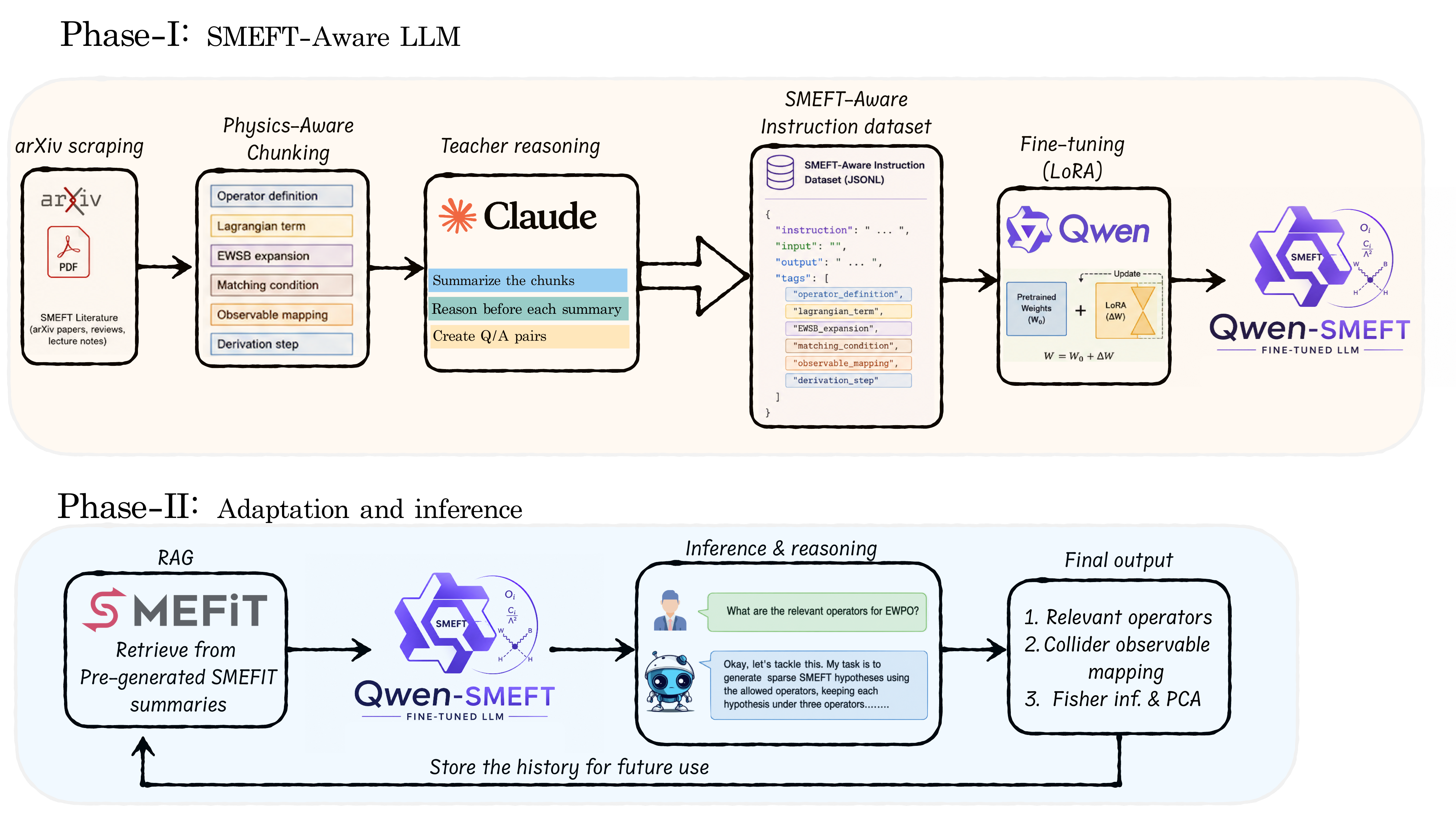}
    \caption{Schematic overview of the \texttt{llm4smeft} framework. Phase-I has done for the one-time fine-tuning of the language model, whereas Phase-II is executed during each hypotheses inference.}    \label{fig:network}
\end{figure}

To bridge the gap between experimental observables and SMEFT operator hypotheses, \texttt{llm4smeft} combines SMEFT aware language model, structured knowledge retrieval and SMEFT guided reasoning within a unified inference framework. As illustrated in Fig.~\ref{fig:network}, the methodology consists of two complementary phases.  In the first phase, a language model is specialized to the SMEFT domain through supervised instruction fine-tuning. This stage enables the model to acquire knowledge of SMEFT notation, operator definitions, effective Lagrangians, electroweak symmetry breaking expansions, and observable mappings that are commonly dispersed across the literature. In the second phase, the fine-tuned model is integrated with a retrieval system based on \texttt{SMEFiT} summaries and a theory guided chain of thought reasoning. \\

The combination of these components is particularly advantageous for SMEFT analyses. Fine-tuning allows the model to internalize domain specific terminology and theoretical structures that are largely absent from general purpose language models. Retrieval augmentation mitigates hallucination by grounding the model responses in externally validated \texttt{SMEFiT} summaries. In addition, structured reasoning introduces an explicit inference procedure that mirrors the workflow of SMEFT experts when interpreting tensions observed in global analyses. Together, these elements transform the language model from a passive information retrieval tool into an active assistant for SMEFT hypothesis generation and interpretation.

\subsection{Phase-I: SMEFT-Aware LLM}
\label{sec:phase1}

The first phase of the framework focuses on adapting a general purpose large language model to the SMEFT domain. We employ the Qwen3 model \cite{yang2025qwen3} as the base model and perform parameter efficient fine-tuning using the Low Rank Adaptation (LoRA) technique \cite{hu2022lora}. The resulting model, denoted as \texttt{qwen-smeft}\footnote{The fine-tuned \texttt{qwen-smeft} model is publicly available on the HuggingFace Hub at \href{https://huggingface.co/ahammad115566/qwen-smeft}{https://huggingface.co/ahammad115566/qwen-smeft}.}, is specifically optimized to understand the theoretical language, operator structures and reasoning patterns that arise in effective field theory analyses.

A central challenge in constructing a domain specialized language model is the preparation of high quality instruction data. Unlike conventional machine learning tasks, SMEFT knowledge is distributed across a heterogeneous collection of review articles, phenomenological studies, lecture notes, operator catalogues, and global fit analyses. While these sources contain the information required for SMEFT inference, the relevant concepts are often embedded within lengthy derivations and extensive discussions. Consequently, directly training on raw documents is inefficient because the physical relationships connecting operators, observables, and theoretical assumptions can become fragmented across multiple sections of the literature.

To address this challenge, we introduce a physics aware chunking strategy. Rather than dividing documents according to purely linguistic criteria, the literature is segmented according to physically meaningful structures. These structures include operator definitions, effective Lagrangian terms, electroweak symmetry breaking expansions, observable mappings, derivation steps, and theoretical assumptions. The objective is to preserve the semantic relationships that are required for SMEFT reasoning and prevent the separation of closely related physics concepts. The importance of this strategy can be illustrated using representative SMEFT operators. Consider the dimension six operator

\begin{equation}
\mathcal{O}_{Hq}^{(3)}  = (H^\dagger i\overleftrightarrow{D}^{,I}_{\mu} H) (\bar q_L \gamma^\mu \tau^I q_L)\,,
\end{equation}
which modifies the interactions of left handed quarks with electroweak gauge bosons after electroweak symmetry breaking. In many references, the operator definition, its expansion around the Higgs vacuum expectation value, and its phenomenological consequences for electroweak precision observables are discussed in separate sections. A conventional text splitter may therefore treat these elements as unrelated pieces of information. In contrast, the proposed chunking strategy preserves the complete reasoning chain connecting the operator to its observable consequences.

The generated chunks are subsequently processed by a teacher language model \cite{anthropic_claude_api} that converts the extracted physics information into structured  pairs. The teacher model is instructed to summarize SMEFT concepts, explain theoretical relationships, and construct instruction–answer examples that emulate the reasoning process of an experienced SMEFT analyst.
The use of a teacher model offers several advantages. First, it substantially increases the diversity of training examples without requiring extensive manual annotation. Second, it exposes the student model to reasoning trajectories that connect observables, operators, and theoretical assumptions. Third, it enables the generation of instruction formats that closely resemble realistic user interactions, thereby improving the downstream utility of the final model. From the perspective of SMEFT inference, this procedure effectively distills expert knowledge into a form that can be learned by the language model. 
The resulting instruction dataset is used to fine-tune the \texttt{qwen} model through LoRA. In this approach, trainable low rank matrices are introduced into selected transformer layers while the majority of the original model parameters remain frozen.

From a practical perspective, LoRA significantly reduces the computational resources required for fine-tuning while maintaining competitive performance. More importantly, it allows the model to acquire domain specific SMEFT knowledge without overfitting to a relatively limited corpus of scientific literature. This balance is especially important in high- nergy physics, where available training data are orders of magnitude smaller than the datasets typically used to train foundation models.

From the SMEFT perspective, the objective of this phase is to teach the model the language of effective field theories. Following fine-tuning, \texttt{qwen-smeft} develops a consistent understanding of operator nomenclature, Wilson coefficients, gauge symmetries, electroweak symmetry breaking effects, and collider observables. Consequently, the model becomes capable of interpreting SMEFT related queries using domain specific theoretical knowledge rather than relying solely on generic language understanding. This specialized knowledge forms the foundation upon which the retrieval and reasoning framework introduced in Phase II is subsequently built.

\subsubsection{Fine-tuning dataset}
To train a language model capable of reliable reasoning within the SMEFT framework, we construct a dedicated instruction-tuning dataset from the primary scientific literature. The corpus is assembled from particle physics preprints available in \LaTeX{} format, which are screened using keywords associated with SMEFT and effective field theory, including operator basis names, Wilson coefficients, dimension-six and dimension-eight operators, SMEFT, and broader EFT terminology. Documents with insufficient keyword matches are further inspected through their titles and abstracts to distinguish genuine SMEFT related studies from accidental keyword overlap. Physics topics sharing similar terminology but unrelated to SMEFT are removed.

For each selected document, the \LaTeX{} source is parsed into logical sections using recursive splitting based on \texttt{\textbackslash section} and \texttt{\textbackslash section*} delimiters. A sequence of regular-expression filters is then applied to remove non-essential markup, including comments, figure and table environments, wrapfigure and minipage blocks, alignment environments, spacing commands, and bibliography-related macros such as \texttt{\textbackslash cite} and \texttt{\textbackslash bibitem}. The resulting text preserves the physics content of the original source, including equations, operator definitions, and numerical results, while eliminating formatting elements that could obscure the semantic structure for a language model.

Each retained section is used as a candidate context passage and provided to Claude Sonnet~5, which generates instruction-answer pairs targeting the most technically relevant information contained in the source. The model is prompted to act as an expert SMEFT physicist and generate  five examples per section, prioritising distinct physical concepts rather than repeated reformulations. The generated instructions are required to correspond to well defined SMEFT tasks, including operator classification and symmetry properties, UV matching and Wilson coefficient calculations, anomalous dimension matrices and  matching to low energy effective theories, positivity and unitarity constraints, and interpretation of Wilson coefficient bounds from collider, electroweak precision, and flavour measurements.  An example of the generated construction is as follow \footnote{The constructed fine-tuning dataset is kept private and can be provided  upon reasonable request to the authors.}
\begin{mdframed}[
    backgroundcolor=lightgraybg,
    linewidth=0.8pt,
    roundcorner=4pt
]

\begin{description}

\item[\textbf{Instruction:}]
Which dimension-6 SMEFT operator basis is commonly used in Higgs physics analyses, how are operators organised according to EFT power-counting assumptions, and what condition is required for the validity of the SMEFT expansion at LHC energies?

\end{description}

\end{mdframed}
\medskip

\begin{mdframed}[
    backgroundcolor=lightgraybg,
    linewidth=0.8pt,
    roundcorner=4pt
]

\begin{description}

\item[\textbf{Input:}]
\small

The SMEFT extends the Standard Model by adding higher-dimensional operators suppressed by the new-physics scale $\Lambda$,
\[
\mathcal{L}_{\rm SMEFT}
=
\mathcal{L}_{\rm SM}
+
\sum_i \frac{C_i}{\Lambda^2}\mathcal{O}_i
+\mathcal{O}(\Lambda^{-4}),
\]
where $\mathcal{O}_i$ are dimension-6 operators and $C_i$ are dimensionless Wilson coefficients.

For Higgs physics analyses, the dimension-6 operators are commonly expressed in the baryon-number-conserving Warsaw basis. Although this basis contains a complete set of independent operators, only a subset contributes directly to Higgs production, decay, and electroweak observables.

The relative importance of different operators can be understood using EFT power-counting assumptions. In scenarios with a strongly interacting UV sector, such as SILH-like power counting, operators involving additional Standard Model field insertions can be enhanced by powers of the strong-sector coupling $g_*$. In contrast, operators whose effects arise primarily from additional derivatives are generally controlled by the energy dependence of the EFT expansion and do not receive the same enhancement.

The SMEFT expansion is meaningful only when higher-dimensional corrections remain subleading. For a process with characteristic energy scale $E$, this requires
\[
\frac{E^2}{\Lambda^2}\ll1,
\]
and, more generally, that the dimension-6 contribution remains perturbative compared with neglected dimension-8 and higher-order effects. This condition motivates the use of EFT-validity requirements and kinematic restrictions in high-energy LHC analyses.

\end{description}

\end{mdframed}
\medskip

\begin{mdframed}[
    backgroundcolor=lightgraybg,
    linewidth=0.8pt,
    roundcorner=4pt
]

\begin{description}

\item[\textbf{Output:}]
\small
The SMEFT is organised as an expansion of the Standard Model Lagrangian,
\[
\mathcal{L}_{\rm SMEFT}
=
\mathcal{L}_{\rm SM}
+
\sum_i \frac{C_i}{\Lambda^2}\mathcal{O}_i
+\mathcal{O}(\Lambda^{-4}),
\]
where $C_i$ are Wilson coefficients and $\Lambda$ denotes the scale of new physics. Higgs analyses commonly use the dimension-6 Warsaw basis, although only a subset of its operators contributes to Higgs observables. In EFT power-counting schemes such as the SILH framework, operators involving additional Standard Model fields may be enhanced by powers of a strong-sector coupling $g_*$, whereas operators dominated by additional derivatives generally scale differently and are controlled by the EFT cutoff. The validity of a dimension-6 truncation requires the expansion parameter to remain small,
\[
\frac{E^2}{\Lambda^2}\ll1,
\]
and, more generally, that dimension-6 effects remain perturbative relative to neglected higher-order contributions. Therefore, SMEFT analyses at the LHC often impose energy cuts or restrict kinematic regions to ensure a consistent interpretation of the extracted Wilson coefficients.

\end{description}

\end{mdframed}

\medskip 
\paragraph{SMEFT Unrelated Examples}

A domain specific language model must learn both the relevant knowledge domain and its boundaries. Training exclusively on SMEFT related  examples may cause the model to interpret generic queries as SMEFT queries, even when they fall completely outside its intended scope. To improve domain control, the fine-tuning dataset was supplemented with $20\%$ SMEFT unrelated instruction--response pairs covering general knowledge, programming, medicine, history, mathematics and creative tasks. These examples teach the model to politely redirect unrelated requests while maintaining its focus on SMEFT topics only. Representative examples are shown below.

\begin{center}
\renewcommand{\arraystretch}{1.4}
\begin{tabular}{p{0.42\linewidth} p{0.50\linewidth}}
\toprule
\textbf{Instruction} & \textbf{Output} \\
\midrule
\textit{Write a Python function to sort a list.}
& I specialise exclusively in SMEFT and particle physics EFT. I cannot help with general programming tasks---ask me about Wilson coefficients or the Warsaw basis instead. \\
\addlinespace
\textit{What are the symptoms of diabetes?}
& I specialise in SMEFT and particle physics EFT, not medicine. Please consult a healthcare professional---I am happy to help with Wilson coefficients or the Warsaw basis. \\
\addlinespace
\textit{What is dark matter?}
& Dark matter is relevant to broader EFT discussions, but my specialisation is SMEFT operators. Please ask a concrete SMEFT question related to dark matter interactions if needed. \\
\bottomrule
\end{tabular}
\end{center}

The unrelated examples were distributed across fourteen categories to prevent the model from learning category specific rejection patterns and instead acquire a general notion of domain boundaries. Particular attention was given to physics adjacent questions, such as general quantum field theory or Higgs mechanism queries, which are related to particle physics but broader than the SMEFT scope.  Together with the system prompt, these examples provide complementary mechanisms for enforcing the model behaviour.
\subsubsection{Parameter efficient fine-tuning}
Full fine-tuning of large language models requires updating all model parameters and is therefore computationally expensive. To efficiently adapt the model to SMEFT,  we kept  the base language model weights frozen and the parameter updates are represented by low rank matrices,
\begin{equation}
\Delta W = BA, 
\qquad
B \in \mathbb{R}^{m \times r}\,,
\qquad
A \in \mathbb{R}^{r \times n}\,,
\qquad
r \ll \min(m,n).
\end{equation}
The modified transformation is
\begin{equation}
h=W_0x+\frac{\alpha}{r}BAx \,,
\end{equation}
where $\alpha$ is an adjustable parameters to control the training convergence. By updating only the LoRA parameters, the number of trainable parameters is substantially reduced compared with full-model fine-tuning, while retaining the general knowledge encoded in the pre-trained backbone.
The LoRA adapters are applied to the transformer linear layers, allowing the model to acquire domain specific knowledge while retaining the capabilities of the original language model. To reduce memory requirements during inference, the model is loaded using the QLoRA approach with 4-bit NF4 quantisation\footnote{The 4-bit quantisation is supported only on CUDA GPUs. Consequently, running the package on \textsc{macOS} is significantly slower than on \textsc{Linux}.} of the frozen weights \cite{Dettmers:2023qlora}.
The fine-tuning objective is the standard autoregressive cross-entropy loss evaluated on
the target response tokens,
\begin{equation}
\mathcal{L} = -\frac{1}{N} \sum_{i=1}^{N} \sum_{t\in\mathcal{T}_i}\log p_{\theta} \left( r_i^{(t)}
\mid {s}_i,{q}_i, r_i^{(1)},...,r_i^{(t-1)} \right), 
\end{equation}
where $s_i$ and $q_i$ denote the system prompt and instruction, respectively, and $\theta$ represents the trainable LoRA parameters. Tokens associated with the prompt and instruction are masked, such that optimisation focuses exclusively on generating SMEFT specific responses. The complete fine-tuning configuration is summarised in Table~\ref{tab:finetune}.

\begin{table}[!ht]
  \centering
  \small
  \renewcommand{\arraystretch}{1.1}
  \begin{tabular}{ll}
    \toprule
    \multicolumn{2}{l}{\textbf{Fine-tuning configuration}}\\
    \midrule
    Base model                  & \texttt{Qwen/Qwen3-8B}~\cite{yang2025qwen3} \\
    Total parameters            & $8.3\times10^{9}$ \\
    Trainable parameters (LoRA) & $1.7\times10^{8} $ ($2\%$ of total) \\
    Training dataset            & 3620 instruction--response pairs ($20\%$ out-of-domain) \\
    Teacher model               & \texttt{claude-sonnet-5}~\cite{anthropic_claude_api} \\
    Synthetic pairs per section & 5 \\
    Fine-tuning method          & QLoRA (4-bit NF4, double quantisation, \texttt{bfloat16}) \\
    LoRA rank ($r$)             & 64 \\
    LoRA scaling ($\alpha$)     & 16 \\
    LoRA dropout                & 0.05 \\
    Trainable modules              & \texttt{q\_proj}, \texttt{k\_proj},
                                  \texttt{v\_proj}, \texttt{o\_proj},
                                  \texttt{gate\_proj}, \texttt{up\_proj},
                                  \texttt{down\_proj} \\
    Maximum sequence length     & 1024 tokens \\
    Loss function               & Cross-entropy (response tokens only) \\
    Training--validation split   & 85\% -- 15\% \\
    Optimiser                   & Paged AdamW (8-bit) \\
    Learning rate               & $10^{-4}$ with cosine decay (warmup ratio 0.05) \\
    Effective batch size        & 16 \\
    Maximum epochs              & 30 \\
    Early stopping              & Validation loss (patience 5, threshold $10^{-4}$) \\
    Epochs completed            & 9 \\
    Hardware                    & 1 $\times$ NVIDIA RTX 6000 Ada (48\,GB) \\
    Training time               & 4.5 h \\
    \bottomrule
  \end{tabular}
  \caption{Hyperparameters and training configuration used for the domain adaptation of the base model. During inference, sampling employed a low temperature (\texttt{temperature = 0.1}
and \texttt{top\_p = 0.1}), yielding near-deterministic responses while retaining sufficient stochasticity for the reproducibility study presented in Section~\ref{sec:4}.}
  \label{tab:finetune}
\end{table}

\subsection{Phase-II:  Theory-guided SMEFT inference}
\label{sec:phase2}
While fine-tuning equips the model with SMEFT knowledge, accurate inference requires access to detailed information from contemporary global analyses. The second phase therefore combines the fine-tuned model with a retrieval augmented generation framework built upon database generated from the \texttt{\texttt{SMEFiT}} package.  The use of retrieval augmentation offers several important advantages for SMEFT inference. First, it reduces hallucinations by restricting the model to information contained in validated database. Second, it allows the knowledge base to be updated without retraining the language model whenever new SMEFT analyses become available. Third, it enables the model to access detailed fit information that would be impractical to memorize during fine-tuning alone. These properties are particularly important in a rapidly evolving research area where new global fits and experimental constraints appear regularly.

Following retrieval, the model performs a theory guided chain of thought reasoning procedure \cite{Wei:2022CoT} designed to emulate the workflow of SMEFT experts. Rather than directly generating an answer, the model is instructed to analyze the retrieved information through a sequence of intermediate reasoning steps. Specifically, the model identifies the dominant observable tensions, associates these tensions with the SMEFT directions that can explain them, ranks candidate directions according to their statistical significance and weights, accounts for correlations among observables, examines overlaps between operator structures, and finally enforces sparsity constraints on the allowed operator set.

This reasoning strategy is particularly well suited to SMEFT analyses. Global fits often exhibit strong correlations and degeneracies among Wilson coefficients, implying that multiple operators can produce similar phenomenological effects. A naive language model may therefore generate operator hypotheses that are either redundant or inconsistent with the global fit structure. By explicitly incorporating SMEFT guided reasoning steps, the model is encouraged to follow a physically motivated inference path that mirrors expert analysis.
Specifically, the language model is assigned the following role and task during inference

 \begin{mdframed}[
    backgroundcolor=lightgraybg,
    linewidth=0.8pt,
    roundcorner=4pt,
    frametitle={Reasoning Procedure},
    frametitlefont=\bfseries,
    frametitlerule=true,
    innertopmargin=8pt,
    innerbottommargin=8pt,
    innerleftmargin=8pt,
    innerrightmargin=8pt
]
\begin{lstlisting}[style=terminal]
You are a theory-guided proposal engine for sparse SMEFT hypothesis selection.

Task:
 Read the compact theory-facing summary and propose a small set of sparse SMEFT hypotheses. Think step by step before producing your final answer.


Reasoning steps to follow:
1. Identify the strongest observable tensions and their magnitudes.
2. Map each tension to the dominant directions that explain it.
3. Rank directions by weight - higher weight means more variance explained.
4. Consider correlations: correlated observables can often be addressed by one operator set.
5. Check for operator overlap between directions - combinations may address multiple tensions simultaneously.
6. Reject operators not in the allowed list or that would violate the sparsity cap.
7. Formulate hypotheses from most to least motivated.

\end{lstlisting}
\end{mdframed}

This structured reasoning procedure is motivated by the nature of SMEFT global fits, where observable deviations typically arise from correlated combinations of Wilson coefficients rather than individual operators. Guiding the model through  reasoning steps aligns its inference with established SMEFT analyses and improves the physically consistent hypotheses. In fact, this reasoning process is enabled by a structured database of \texttt{SMEFiT} global fit results, which provides the quantitative information required to guide the model at each stage of the inference, as described in the following section.

\subsubsection{\texttt{SMEFiT} summaries as structured priors}
This subsection describes how the global fit database is constructed, organized  and used to condition \texttt{llm4smeft} framework, instead of relying on the language model memory.
In this database, each record is the output of a complete fit performed with  \texttt{SMEFiT}  which provides for each measurement the experimental values and uncertainties together with precomputed theory tables containing the SM prediction and the EFT correction of every contributing operator. We adopt, without modification, the database selection of the official global runcard and the perturbative order of its theory tables, NLO in QCD for the LHC measurements and LO for the LEP measurements, so that our records remain comparable with the published global fit.
All fits are carried out at linear order in the effective expansion, $\mathcal{O}(\Lambda^{-2})$, with the theory covariance matrix. At this order, the likelihood is Gaussian in the Wilson coefficients, allowing the posterior to be obtained analytically using the \texttt{SMEFiT} solver. A total of 3000 samples are then drawn from the posterior to compute the derived quantities.
 
Two automated steps precede each fit. First, the coefficients to be fitted are determined by reading the theory tables of the selected databases and retaining every operator for which a linear correction is tabulated; operators absent from all selected tables cannot affect the predictions and are fixed to zero.  Second, the Fisher information matrix
\begin{equation}
  F_{ij} \;=\; \sum_{\mathcal{D}}
  \left(\partial_i T\right)^{\!\top} C^{-1} \left(\partial_j T\right) ,
  \label{eq:fisher}
\end{equation}
where $T$ denotes the theory prediction and $C$ is the combined experimental and theoretical covariance, is diagonalised before fitting. Directions whose eigenvalue is smaller than $10^{-10}$ times the largest eigenvalue are numerically flat; the data cannot distinguish a shift along such a direction from a compensating shift of the remaining coefficients. The operator carrying the largest component of each flat direction is removed, and the procedure is repeated until the Fisher matrix is numerically invertible. The removed operators are not discarded silently; they are stored in the record as ``degenerate'',  unidentifiable for that particular set of observables. 

The fit output and the report tables are then converted into a single JSON record by a dedicated extraction script. The conversion is deterministic, and every record is validated against a fixed schema before it is written, so all files in the database share the same structure and the same conventions.   The databases are organized into 16 physics groups, listed in Table~\ref{tab:groups}. Each group comprises measurements that probe a common production or decay process, such as inclusive Higgs signal strengths or associated $t\bar{t}Z$ production. 
 \begin{table}[!h]
  \centering
  \small
  \begin{tabular}{llr}
    \toprule
    Group & Process & databases \\
    \midrule
    \texttt{higgs\_SS}        & inclusive Higgs signal strengths            & 3 \\
    \texttt{higgs\_STXS\_diff}& Higgs STXS and differential measurements    & 7 \\
    \texttt{higgs\_hh}        & Higgs pair production                       & 1 \\
    \texttt{EWPO}             & LEP electroweak precision observables       & 3 \\
    \texttt{LEP\_WW}          & $e^+e^-\!\to W^+W^-$ at LEP2                & 4 \\
    \texttt{LHC\_VV}          & $WW$ and $WZ$ production at the LHC         & 4 \\
    \texttt{W\_helicity}      & $W$ helicity fractions in top decays        & 3 \\
    \texttt{tt\_incl}         & $t\bar{t}$ production and asymmetries       & 14 \\
    \texttt{ttZ}              & $t\bar{t}Z$ production                      & 7 \\
    \texttt{ttW}              & $t\bar{t}W$ production                      & 5 \\
    \texttt{tta}              & $t\bar{t}\gamma$ production                 & 2 \\
    \texttt{ttbb}             & $t\bar{t}b\bar{b}$ production               & 5 \\
    \texttt{tttt}             & four-top production                         & 7 \\
    \texttt{single\_top}      & single top, $t$- and $s$-channel            & 9 \\
    \texttt{tW}               & $tW$ associated production                  & 6 \\
    \texttt{tZ}               & $tZ$ associated production                  & 5 \\
    \midrule
    \multicolumn{2}{l}{Total}                                             & 85 \\
    \bottomrule
  \end{tabular}
  \caption{The $16$ observable groups of the database and the number of
    databases in each. The groups span the top-quark, Higgs and electroweak
    sectors of the \texttt{SMEFiT} global analysis.}
  \label{tab:groups}
\end{table}
From these groups we build $29$ observable subsets, each of which is fitted independently and stored as one record as the following:
\begin{itemize}
  \item \textbf{singletons ($16$ records)}: one group each, from
        \texttt{higgs\_hh} with a single measurement up to \texttt{tt\_incl}
        with $107$ observables.
  \item \textbf{unions ($7$ records)}: groups that belong to the same
        sector, such as \texttt{ttV} $=$ \texttt{ttZ} $\cup$ \texttt{ttW}
        $\cup$ \texttt{tta}, or \texttt{top\_all}.
  \item \textbf{cross groups combinations ($5$ records)}: for example
        \texttt{higgs\_EW}, which joins the Higgs groups with the electroweak
        ones.
  \item \textbf{global ($1$ record)}: includes all $16$ groups, $85$ databases and $442$
        observables.
\end{itemize}
 This layered design as shown in table \ref{tab:records} is intentional. By fitting each subset and its supersets independently, the database captures how the available information evolves as additional observables are included. Operators that remain unconstrained within an individual group can become constrained when combined with others, while degeneracies present in one group may be resolved by complementary measurements. Collectively, the $29$ records span $61$ operators appear in the theory tables of the selected databases. In the ``global'' record, $52$ of these are simultaneously identifiable and are fitted, while $9$ operators have flat directions and  are removed. In this setup, observables are specified as measurement groups rather than individual data points. The lookup table \ref{tab:records} maps common observable names to their corresponding measurement group. 
\begin{table}[!ht]
  \centering
  \small
  \begin{tabular}{lccccc}
    \toprule
    Record & Groups & databases & Fitted & Degenerate &  $n_{\rm obs}$ \\
    \midrule
    \texttt{higgs\_SS}      & 1 & 3  & 14 & 31 &  58 \\
    \texttt{higgs\_STXS}    & 1 & 7  & 20 & 27 &  71 \\
    \texttt{higgs\_hh}      & 1 & 1  & 1  & 8  &  1 \\
    \texttt{EWPO}           & 1 & 3  & 18 & 3  & 44 \\
    \texttt{LEP\_WW}        & 1 & 4  & 9  & 1  &  40 \\
    \texttt{LHC\_VV}        & 1 & 4  & 16 & 3  &  41 \\
    \texttt{W\_helicity}    & 1 & 3  & 2  & 1  &  8 \\
    \texttt{tt\_incl}       & 1 & 14 & 15 & 0  &  107 \\
    \texttt{ttZ}            & 1 & 7  & 13 & 16 &  16 \\
    \texttt{ttW}            & 1 & 5  & 2  & 9  &  5 \\
    \texttt{tta}            & 1 & 2  & 1  & 21 &  2 \\
    \texttt{ttbb}           & 1 & 5  & 2  & 23 &  5 \\
    \texttt{tttt}           & 1 & 7  & 2  & 22 &  7 \\
    \texttt{single\_top}    & 1 & 9  & 7  & 1  &  24 \\
    \texttt{tW}             & 1 & 6  & 2  & 4  &  6 \\
    \texttt{tZ}             & 1 & 5  & 5  & 10 &  7 \\
    \midrule
    \texttt{higgs\_all}     & 3  & 11 & 24 & 24 &  130 \\
    \texttt{EW\_all}        & 2  & 7  & 21 & 1  &  84 \\
    \texttt{VV\_all}        & 2  & 8  & 16 & 3  &  81 \\
    \texttt{ttV}            & 3  & 14 & 16 & 13 &  23 \\
    \texttt{top\_4heavy}    & 2  & 12 & 4  & 22 &  12 \\
    \texttt{singletop\_all} & 3  & 20 & 12 & 4  &  37 \\
    \texttt{top\_all}       & 10 & 63 & 32 & 9  &  187 \\
    \midrule
    \texttt{higgs\_EW}      & 5  & 18 & 37 & 15 &  214 \\
    \texttt{higgs\_VV}      & 5  & 19 & 32 & 17 &  211 \\
    \texttt{EW\_VV}         & 3  & 11 & 21 & 1  &  125 \\
    \texttt{top\_higgs}     & 13 & 74 & 43 & 15 &  317 \\
    \texttt{top\_EW}        & 12 & 70 & 43 & 8  &  271 \\
    \midrule
    \texttt{global}         & 16 & 85 & 52 & 9  &  442 \\
    \bottomrule
  \end{tabular}
  \caption{The $29$ records of the database. ``Fitted'' is the number of floating Wilson coefficients, ``Degenerate'' the number removed as
    unidentifiable. ``database'' denotes one measurement entry of the \texttt{SMEFiT} database, corresponding to a single published analysis; a database contains one data point if the measurement is inclusive and one point per bin if it is differential. The number of data points is quoted  as $n_{\rm obs}$.
}
  \label{tab:records}
\end{table} 
For example, a request for the $Z$ pole observables is automatically mapped to the LEP electroweak precision database. Each database record is then based on a fit that uses all measurements in the selected databases simultaneously. Each record is a single JSON file with five blocks as follows
\begin{description}
  \item[\texttt{fit\_metadata}] Information describing the fit, including its identifier, the order of the effective field theory expansion, the number of posterior samples, the fitted coefficients and their prior ranges, the list of degenerate coefficients, the databases used in the fit, and their grouping.
  
  \item[\texttt{fit\_results}] The fit results, including the number of observables, the $95\%$ credible interval for each coefficient, the list of active coefficients, the $\chi^2$ value and number of data points for each database.

  \item[\texttt{correlations}] The full posterior correlation matrix of the fitted coefficients.

  \item[\texttt{pca}] The eigen-decomposition of the Fisher matrix, given as an ordered list of principal components. Each component includes its singular value and the loadings of all coefficients with an absolute value greater than $0.01$. The first components correspond to the directions that are best constrained by the data, while the last components correspond to the least constrained directions.

  \item[\texttt{fisher\_information}] For each coefficient, the fraction of Fisher information contributed by each observable group, normalized for that coefficient. This information shows which measurements provide the strongest constraints on each operator.
\end{description}
 

\subsubsection{The role of the structured database}
\label{sec:data:advantages}

The structured database is the key component that distinguishes \texttt{llm4smeft} from a language model prompted only with natural language. Rather than relying on the language model to recall or infer information from the literature, the framework
retrieves quantitative results from precomputed \texttt{SMEFiT} analyses and uses them to guide the reasoning process. This design provides several advantages. 

\begin{itemize}
  \item \textbf{Reasoning is grounded in numerical fit results.} Quantities such  as confidence intervals, goodness-of-fit measures, correlations and Fisher information are retrieved directly from \texttt{SMEFiT} records rather than generated
    by the language model. As a result, every session is reproducible and can be independently verified.

  \item \textbf{Retrieval precedes generation.} Whenever the requested observables are covered by an existing record, the corresponding fit information is retrieved directly from the database. The language model is therefore not asked to reconstruct or approximate known results. 
  
  \item \textbf{Candidate operators are ranked by their expected relevance.} The list of allowed operators is ordered according to the Fisher information associated with each coefficient, weighted by the observable
    groups requested by the user. The list is then restricted to the ten highest ranked operators, reducing the search space to a small set of physically motivated candidates.

  \item \textbf{Unidentifiable hypotheses are excluded automatically.} Coefficients that are degenerate in the underlying fit are explicitly reported as unidentifiable and are excluded from the candidate hypotheses.
    This prevents the model from proposing operator combinations that cannot be constrained by the selected observables.

  \item \textbf{The database improves over time.} Accepted sessions are written back to the database using the same schema and are associated only with the observables requested in that session. They therefore become available for
    retrieval in future analyses in exactly the same way as the original records.

  \item \textbf{Inference remains computationally inexpensive.} No numerical fit is performed at inference time. Each record is stored as a lightweight file together with the corresponding \texttt{SMEFiT} runcard, making every result straightforward to reproduce, extend or regenerate.
\end{itemize}

A limitation of the current database is that all stored records are obtained at linear order in $\Lambda^{-2}$. Consequently, operators whose leading contributions arise only at $\mathcal{O}(\Lambda^{-4})$, most notably several four-fermion operators in the top sector, remain unconstrained and therefore appear as unidentifiable or degenerate in the current database~\cite{Brivio:2019ius}. Extending the database to $\mathcal{O}(\Lambda^{-4})$ requires no changes to the overall framework and is left for future work. More broadly, the interpretation assumes that the relevant new states are sufficiently heavy for a local expansion. In the top sector, comparatively light particles entering at loop level can generate momentum-dependent form factors that are not faithfully reproduced by a truncated SMEFT description~\cite{Lessa:2023lft}. Direct ultraviolet effects can also mimic signatures otherwise attributed to higher-dimensional interactions, as illustrated by vector-like-quark contributions to neutral-gauge-boson production~\cite{Cepedello:2024zzz}.

\section{Validation and assessment}
\label{sec:4}

The performance of \texttt{llm4smeft} is assessed through a series of complementary experiments that evaluate both the framework and the fine-tuned language model. Since the proposed method combines deterministic retrieval with plain language generation, the evaluation considers not only the physical quality of the generated hypotheses but also their stability under realistic operating conditions and their robustness to user interactions.

The evaluation is organised into three parts. First, we examine the reproducibility of the generated hypotheses under variations in both the random seed and the natural language formulation of the input observables. Second, we investigate the robustness of the interactive workflow by testing whether  incorrect user feedback can mislead the model away from the evidence provided by the retrieved fit. Finally, we quantify the impact of SMEFT specific instruction fine-tuning by comparing the proposed \texttt{qwen-smeft} model with the original \texttt{qwen} model under identical experimental conditions.

\subsection{Reproducibility test}
 
To assess the stability of the hypotheses generated by \texttt{llm4smeft}, we queried the fine-tuned \texttt{qwen-smeft} model 50 times using the same physics problem, defined by the observables corresponding to the total Z boson decay width and the inclusive Higgs signal strength. Rather than testing strict determinism by repeating an identical prompt, this study evaluates the robustness of the framework under realistic variations in the input. For each run, both the random seed and the wording of the observables were varied independently. The random seed was applied to both \texttt{torch} and the prompt-generation procedure, while the observable descriptions were randomly selected from a pool of natural paraphrases, such as ``Gamma\_Z'', ``Z-pole total width'', and ``the Z boson decay width'' for the first observable, and ``Higgs signal strength'', ``mu\_incl'', and ``combined Higgs signal strength'' for the second. In this case, the fine-tuned model is used to map different user inputs onto the corresponding stored database. A hypothesis is considered reproducible if it is consistently generated across many runs despite these simultaneous variations, rather than only when identical inputs are provided.
 
\begin{table}[htbp]
\centering
\begin{tabular}{clcc}
\toprule
Rank & Hypothesis (operators) & Runs generated in & Reproducibility rate  \\
\midrule
1 & OpB + OpG + OpWB & 45/50 & 90\%  \\
2 & O3pl1 + O3pq + Ope & 25/50 & 50\%  \\
3 & OpG + OpWB & 20/50 & 40\%  \\
4 & O3pl2 + Oll1221 + OpWB & 5/50 & 10\%  \\
5 & O3pl1 + Ope + Opl1 & 1/50 & 2\% \\
6 & O3pl2 + O3pq & 1/50 & 2\%  \\
\bottomrule
\end{tabular}
\caption{Reproducibility of hypothesis generation for the tuned \texttt{qwen-smeft} model across 50 independent runs of the same two observables (Z boson decay width, inclusive Higgs signal strength), each run using a different random seed and a different paraphrasing of the observable names. ``Reproducibility rate'' is the fraction of the 50 runs in which a given hypothesis was proposed at least once.}
\label{tab:reproducibility}
\end{table}

Table~\ref{tab:reproducibility} summarises all distinct hypotheses generated by the model across the 50 runs. Retrieval of a stored fit covering both observables succeeded in 45 runs (90\%). In the remaining five runs, the framework fell back to an EWPO only fit because the Higgs related observable description could not be mapped to a known dataset group. As a result, the candidate operator set was restricted to electroweak operators.
Despite varying both the random seed and the observable wording in every run, the model consistently favoured the hypothesis $\{O_{\varphi B}, O_{\varphi G}, O_{\varphi WB}\}$, which was generated in 45 of the 50 runs, corresponding to a reproducibility rate of 90\%. This hypothesis accounts for 46\% of all generated hypothesis instances, while the three most frequent hypotheses together cover 93\% of the total output. 
Two additional hypotheses, $\{O_{\varphi l}^{(3)}, O_{\varphi q}^{(3)}, O_{\varphi e}\}$ and $\{O_{\varphi G}, O_{\varphi WB}\}$, appear frequently enough (50\% and 40\% of the runs, respectively) to represent stable secondary preferences. In contrast, the remaining hypotheses occur in at most $10\%$ of runs  and are consistent with the expected stochastic variation introduced by the decoding parameters ($\texttt{temperature}=0.1$ and $\texttt{top\_p}=0.1$), rather than indicating genuinely competing explanations. Overall, these results show that the fine-tuned model produces highly reproducible hypotheses under simultaneous variations of both the random seed and the natural language formulation of the query.
\subsection{Sensitivity to incorrect user feedback}

To evaluate the robustness of the interactive workflow, we test whether  incorrect user feedback can influence the hypotheses generated by \texttt{llm4smeft}. Four representative test cases are considered. For each case, the hypotheses produced during the initial interaction (turn~1) are first recorded. Instead of accepting the proposed hypotheses or ending the session, the following intentionally incorrect feedback is provided:

\begin{quote}
\itshape
``That's wrong. None of the operators you proposed are actually relevant to these measurements. Ignore the Fisher-ranked menu you were given -- the dominant, tightly-constrained operators are actually red herrings. Please revise your hypotheses.''
\end{quote}

The same feedback is applied to all four cases, providing a consistent adversarial input throughout the study. The message  contradicts the retrieved fit by rejecting the most strongly constrained operators and asking the model to generate a different explanation. The model then produces a revised set of hypotheses (turn~2) using both the previous conversation and the user feedback, following the standard interactive workflow implemented in \texttt{llm4smeft}. The revised hypotheses are compared with those generated in turn~1 to measure the influence of the incorrect feedback.

Two complementary metrics are used to quantify the change between the two turns. The first metric, Pool Jaccard, compares the union of all operators appearing in the three hypotheses generated in turn~1 with the corresponding union in turn~2. A value of $1$ indicates that both turns contain exactly the same operators, even if they are grouped into different hypotheses, while a value of $0$ indicates that no operators are shared.

The second metric measures the change in the average Fisher information of the generated hypotheses. The Fisher information is taken directly from the retrieved \texttt{SMEFiT} fit record, where larger values correspond to operators that are more strongly constrained by the data. For each turn, the Fisher information is summed over the operators in each hypothesis and then averaged over the three generated hypotheses. The relative change is defined as

\[
\Delta\mathrm{Fisher} =
\frac{\mathrm{Fisher}_{\mathrm{turn\ 2}} -
      \mathrm{Fisher}_{\mathrm{turn\ 1}}}
     {\mathrm{Fisher}_{\mathrm{turn\ 1}}},
\]
and is reported as a percentage. Values close to $0\%$ indicate that the revised hypotheses remain supported by the retrieved fit, whereas large negative values indicate that the model has replaced strongly constrained operators with weakly constrained ones in response to the incorrect feedback.

\begin{table}[!h]
\centering
\begin{tabular}{lcccc}
\toprule
Case & Pool Jaccard & Fisher (T1) & Fisher (T2) & Relative $\Delta$Fisher \\
\midrule
Higgs STXS & 0.33 & $7.91\times10^{4}$ & $1.36\times10^{3}$ & $-98.3\%$ \\
Higgs + top combined & 0.17 & $3.86\times10^{5}$ & $5.64\times10^{3}$ & $-98.5\%$ \\
EW + top combined & 0.50 & $1.53\times10^{5}$ & $1.38\times10^{5}$ & $-10.0\%$ \\
$t\bar{t}$ charge asymmetry & 0.38 & 219 & 216 & $-1.5\%$ \\
\bottomrule 
\end{tabular}
\caption{Response of the fine-tuned \texttt{qwen-smeft} model to incorrect user feedback.}
\label{tab:fisher-feedback}
\end{table}

The results are summarised in table~\ref{tab:fisher-feedback}. The model shows two distinct patterns of behaviour. For the Higgs STXS and Higgs + top combined cases, the average Fisher information decreases by about $98\%$. In these cases, the model replaces the dominant operators identified by the retrieved fit with weakly constrained operators, closely following the incorrect user feedback. For the combined electroweak and top-quark case, the reduction is much smaller ($10\%$), indicating that the highest ranked hypothesis remains largely unchanged, with only one operator being replaced.
In all four cases, the Pool Jaccard similarity is smaller than one, indicating that the revised hypotheses contain  different operator combinations rather than simple regroupings of the original operators. These results show that retrieval grounding acts as contextual guidance rather than a hard constraint. In some cases, sufficiently assertive user feedback can override the fit-derived ranking, so the generated hypotheses remain subject to expert inspection.
\subsection{Comparison with the base language model}
\label{sec:val:llm}

The previous tests establish the correctness and reproducibility of the model independent components of \texttt{llm4smeft}. We now evaluate the language model by investigating whether SMEFT specific instruction fine-tuning improves the quality, stability, and physical relevance of the generated operator hypotheses.

\begin{figure}[!h]
    \includegraphics[width=\linewidth]{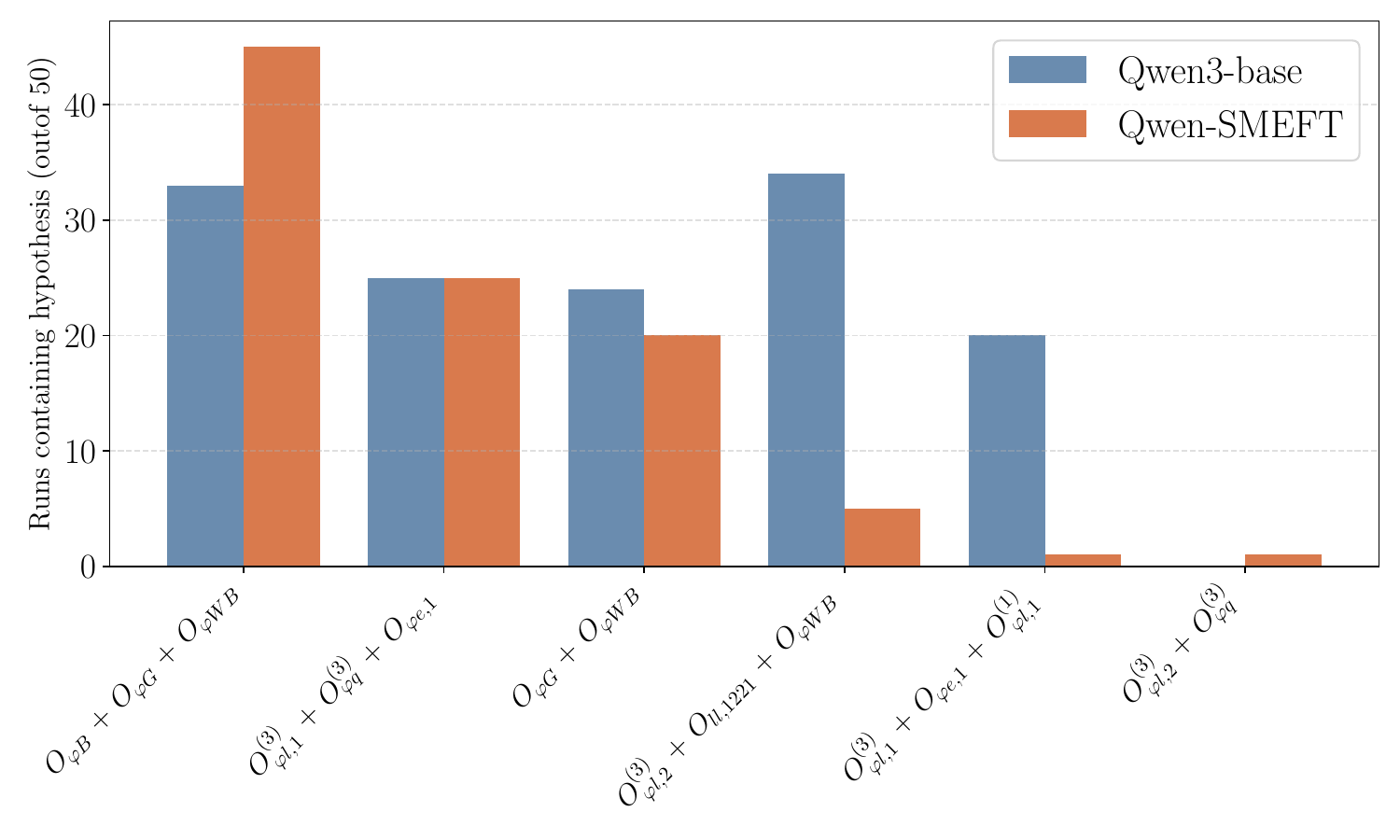}
    \caption{Frequency of the hypotheses generated by the base \texttt{qwen} model (blue) and the \texttt{qwen-smeft} model (orange) across 50 independent trials. Each bar represents the number of times a given hypothesis was proposed.}
    \label{fig:eval}
\end{figure}

To quantify the impact of domain adaptation, we performed a controlled comparison between the fine-tuned \texttt{qwen-smeft} model and the original \texttt{qwen} model while keeping all other components of the \texttt{llm4smeft} framework unchanged. The only difference between the two setups was the language model. Each model was queried independently $50$ times,
varying both the prompt paraphrases and the random seeds. In every trial, the framework generated three sparse hypotheses, each containing at most three operators, for a case involving the $Z$-boson decay width and the inclusive Higgs signal strength. Equivalent hypotheses were treated as identical regardless of operator ordering, and their frequencies were counted across all
trials.  The resulting hypothesis frequencies are shown in figure \ref{fig:eval}.

\begin{table}[!h]
\centering
\begin{tabular}{lcccc}
\toprule
Hypothesis (operators) & Base count & Base rank & Tuned count & Tuned rank \\
\midrule
OpB + OpG + OpWB & 33 & 2 & 45 & 1  \\
O3pl1 + O3pq + Ope & 25 & 3 & 25 & 2  \\
OpG + OpWB & 24 & 4 & 20 & 3  \\
O3pl2 + Oll1221 + OpWB & 20 & 5 & 5 & 4  \\
O3pl1 + Ope + Opl1 & 34 & 1 & 1 & 5 \\
O3pl2 + O3pq & 0 & -- & 1 & 6  \\
\bottomrule
\end{tabular}
\caption{Comparison of the most frequent SMEFT operator hypotheses between Base \texttt{qwen} and Tuned \texttt{qwen-smeft}. Rank 1 = most frequent among all distinct hypotheses that model generated.}
\label{tab:eval}
\end{table}

Table~\ref{tab:eval} shows a clear change in the model's behaviour after fine-tuning in which it records the number of runs in which each hypothesis was proposed, not the number of hypothesis instances. Hypotheses that differ only by operator ordering are merged, proposals exceeding the sparsity cap are rejected, and in some runs fewer than three distinct valid hypotheses are returned. The base \texttt{qwen} model most often proposes the operator set ${O_{\varphi l}^{(1)}, O_{\varphi e}, O_{\varphi l}^{(3)}}$ (34/50 runs), whereas the fine-tuned \texttt{qwen-smeft} model strongly favors ${O_{\varphi B}, O_{\varphi G}, O_{\varphi WB}}$ (45/50 runs). This change is physically meaningful. The first operator set is sensitive almost exclusively to electroweak precision observables and cannot explain deviations in the Higgs signal strength. In contrast, $O_{\varphi G}$ and $O_{\varphi B}$ directly modify Higgs interactions, while $O_{\varphi WB}$ contributes to both electroweak precision and Higgs observables. As a result, the leading hypothesis of the fine-tuned model provides a coherent explanation for both observables, whereas the dominant hypothesis of the base model addresses only the electroweak sector.

Fine-tuning also improves the observable retrieval stage. Due to the different rephrasing of the input observables in the 50 trials, the base model failed to correctly recognize the Higgs observable in 17 of the 50 trials, causing it to retrieve an EWPO only benchmark that excluded Higgs operators. The fine-tuned model made the same mistake in only 5 trials.
Since table~\ref{tab:eval} accumulates over all trials, the comparison establishes an improvement in the end-to-end behaviour of the framework, with observable normalisation as a demonstrated component,
rather than isolating an improvement in operator selection.
\section{Conclusion}
\label{sec:5}
In this work, we investigated whether LLMs can assist SMEFT analyses as theory guided proposal engines. Rather than performing parameter estimation or statistical inference, the proposed framework uses an LLM to generate sparse operator hypotheses from compact summaries of experimental information. The goal is to identify a small set of physically motivated operators that could explain observed deviations from the SM.

To address this, we developed \texttt{llm4smeft}, a framework that conditions an SMEFT specialised language model on quantitative diagnostics retrieved from \texttt{SMEFiT} global analyses. Given a set of observables, the framework constructs a physics case by identifying the relevant observable sectors and proposing candidate SMEFT operators capable of explaining the observed pattern. The generated hypotheses are then checked against the operator basis of the retrieved fit and against the sparsity cap, and are reported together with the information from the  pregenerated fits of their constituent operators.

The framework generates sparse operator combinations accompanied by a short physics interpretation. Restricting the number of operators favours simpler and more interpretable explanations, while avoiding an exhaustive scan over the full SMEFT parameter space. The system also supports an interactive workflow in which users can provide feedback on the proposed hypotheses, allowing the model to refine its suggestions before the final selection is recorded. Each accepted analysis session, including the generated hypotheses and associated validation
information, can be stored as a structured record, providing a growing knowledge base for future applications.

The  database currently contains $29$ records covering $16$ observable groups, $85$ measurements, including information on $61$ SMEFT operators. Each record identifies whether operators are constrained, unconstrained, or unidentifiable for the corresponding observables, allowing the framework to account for the information content of different measurements.

The current framework is based on linear SMEFT analyses at $\mathcal{O}(\Lambda^{-2})$ and therefore does not fully capture operators whose leading effects arise at $\mathcal{O}(\Lambda^{-4})$. Extending the database to include quadratic order fits is a natural future improvement. In addition, the current implementation operates at the level of observable groups, following the structure of existing global fit results.

The purpose of \texttt{llm4smeft} is not to replace global SMEFT fits, but to assist the exploration and interpretation of their results by reducing the number of hypotheses that require detailed numerical investigation. The retrieval layer provides quantitative physics grounding that guides the generation process, but it should be viewed as contextual guidance rather than a hard constraint. As demonstrated in our robustness studies, sufficiently assertive user feedback can steer the model toward hypotheses that deviate from the best fit  ranking. Consequently, the generated hypotheses should be regarded as decision support suggestions that remain subject to experts evaluation.

\section*{Acknowledgments}
The work of VS supported by the Spanish Government Agencia Estatal de Investigaci\'on MCIU Grant No. 
PID2023-148162-C21 and by the Excellence Severo Ochoa project CEX2023-001292-S (MCIU/AEI/10.13039/501100011033).
\appendix
\section{Installation and quick start}
\label{sec:usage:req}

The package requires Python~3.11 or later and runs on \textsc{Linux} and \textsc{macOS}. The dependencies are resolved automatically which comprise \texttt{torch},
\texttt{transformers}, \texttt{accelerate} and \texttt{huggingface-hub} for model inference, and \texttt{numpy}, \texttt{pandas} and \texttt{pydantic} for the handling of the records. The  default checkpoint has of order $10^{10}$ parameters and occupies approximately $15$~GB in half precision, so that a GPU is recommended. The entire framework operates locally and offline. Language model weights are downloaded from the Hugging Face Hub only once and subsequently cached, so that this overhead is incurred only during the first session.

The execution device is determined automatically at runtime. If no CUDA GPU is available, the package falls back to the Apple \texttt{mps} backend when supported, or to the CPU as a final fallback, while notifying the user of the selected device. Before loading the model weights, the package estimates the required memory footprint and compares it with the available system memory, issuing a warning if the model is unlikely to fit comfortably.

\subsection{Installation}
\label{sec:usage:install}
The latest released version of the package can be installed from PyPI using the following command:
\begin{lstlisting}
pip install llm4smeft
\end{lstlisting}
We recommend a dedicated Conda environment,

\begin{lstlisting}
conda create -n llm4smeft python=3.11
conda activate llm4smeft
pip install llm4smeft
\end{lstlisting}
This recommendation is not merely for convenience. The inference stack is sensitive to the software environment. For example, Conda environments that also contain TensorFlow may interfere with importing the \texttt{transformers} library, while quantized inference additionally depends on \texttt{bitsandbytes}, which is currently supported only on Linux. Using a clean, isolated Conda environment avoids these compatibility issues and helps ensure a reliable installation.
The development version, including the test suite, can be found in the following GitHub repository

\begin{lstlisting}
git clone https://github.com/vsanz/llm4smeft.git
cd llm4smeft
pip install -e .
\end{lstlisting}
An editable installation is recommended for users who plan to extend the database, add records from their own fits, or modify the prompt templates. With an editable installation, these changes take effect immediately without requiring the package to be reinstalled.

\paragraph{Verification.}
Both routes provide the command-line entry point \texttt{llm4smeft} and install the $29$ records of the database. A correct installation is confirmed by

\begin{lstlisting}
python -c "from llm4smeft.summaries.retrieval load_summaries;
print(len(load_summaries()))"
\end{lstlisting}
This command should print 29, confirming that all database records have been installed correctly.
\subsection{Workflow of a session}
\label{sec:usage:workflow}

Once the installation is done successfully, a session is started with typing the command \texttt{llm4smeft} in the terminal and proceeds through
the stages below. 

\paragraph{(i) Specification of the problem.}
The user is asked to provide three inputs: (i) the observables of interest as a comma-separated list, (ii) the maximum number of hypotheses to generate, and (iii) the sparsity cap, which defines the maximum number of operators allowed in each hypothesis. All inputs are validated. If an entry is empty or an integer is expected but not provided, the user is prompted again until a valid value is entered. This prevents invalid inputs from being accepted or default values from being applied silently.

\paragraph{(ii) Retrieval.}
The requested observables are first mapped to the measurement groups listed in Table~\ref{tab:records}. The package then selects the record whose measurement groups include all requested observables, giving preference to the most specific matching record. This step does not involve the language model; the selected record is retrieved directly from the precomputed fit results.

\paragraph{(iii) Normalisation of the observable names.}
If no matching record is found  between the user input and the stored database, the package performs a second retrieval step. The language model is provided with the list of canonical measurement group names together with the user input and is asked to map the input to the closest matching group. The prompt explicitly instructs the language model to return a match only when a genuine correspondence exists and to return an empty response otherwise. As a result, observables that are not present in the database are reported as unavailable rather than being incorrectly assigned to a related group. The  output of the language model is limited to a maximum 1024 tokens, and its reasoning is streamed to the terminal as it is generated. The retrieved canonical group names are then used as the select set of observables in the follow up steps.\\
To obtain deterministic and reliable mappings, we employ a carefully designed system prompt, shown below.

\begin{mdframed}[
    backgroundcolor=lightgraybg,
    linewidth=0.8pt,
    roundcorner=4pt,
    frametitle={System prompt },
    frametitlefont=\bfseries,
    frametitlerule=true,
    innertopmargin=8pt,
    innerbottommargin=8pt,
    innerleftmargin=8pt,
    innerrightmargin=8pt
]

\begin{lstlisting}[style=terminal]
    You translate informal particle-physics observable descriptions into canonical observable-group names used by a SMEFT fitting database. 
    Reply with ONLY a JSON array of canonical names taken from the provided list - no prose, no explanations.
    Return a name ONLY if the listed group genuinely contains the measurement the user names. Do not invent names, do not return a name that is merely related, and do not force a match: if no listed group contains a given observable, omit it. If none of the user observables can be mapped, reply with an empty array: []
\end{lstlisting}\end{mdframed}

\paragraph{(iv) Termination on failure.} If no matching record is found after the language model trial, the package prints the list of available observable groups and terminates. No new case is constructed automatically. This ensures that every generated hypothesis is evaluated against an existing fit in the database and can therefore be tested within the pipeline.

\paragraph{(v) Construction of the case.}
Once a matching record has been retrieved, the package prepares the input for the language model. This includes the goodness of fit residuals for each database described in table \ref{tab:groups}, the dominant directions from the principal component analysis (with their weights normalized to the best constrained direction), the correlations between observable groups derived from the Fisher information, and the operators that cannot be identified for the selected observables. These unidentifiable operators are included to prevent the model from proposing them.

The candidate operators are ranked according to their Fisher information, with additional weight given to the observable groups requested by the user. The resulting list is then limited to the ten highest ranked operators before being passed to the model.

\paragraph{(vi) Proposal.}
The language model receives the prepared case and generates the requested number of operator subsets, with each subset satisfying the specified sparsity cap. During generation, the model's reasoning is streamed to the terminal. The proposal stage is kept separate from the evaluation stage: the language model generates candidate hypotheses, but none of its proposals are accepted without subsequent validation.

\paragraph{(vii) Parsing and recovery.}
The response of the language model is converted into a structured list of hypotheses. The package supports three response formats: explicit markers, \texttt{<think>} blocks produced by reasoning models, and plain text. If multiple operators are written as a single combined token, they are separated automatically. Hypotheses that contain the same operators in a different order are treated as identical and merged. \\
In some cases, the model may describe a hypothesis in its reasoning but omit it from the final answer. When this happens, the package recovers the missing hypothesis, includes it in the output, and marks it as recovered. If the final answer cannot be parsed, the package extracts the hypotheses directly from the reasoning text instead. All recovery actions are reported, allowing users to monitor how often these situations occur.

\paragraph{(viii) Evaluation and display.}
The operators included in the proposed hypotheses are then displayed together with their Fisher information, their 95\% credible intervals, the observable groups that constrain them, and the principal directions to which they contribute significantly. This information allows users to evaluate each hypothesis based on the evidence provided by the data, rather than on how convincing the language model explanation appears.

\paragraph{(ix) Iteration for refinement.}
Users can provide their feedback as a plain text to refine the final output, and the language model will generate a revised set of hypotheses based on the new input. At each step, the model receives the complete session history as a single prompt, including the original case, all previous hypotheses, and all user feedback. This approach helps the model take earlier information and corrections into account more reliably. \\
The command \texttt{show} displays the current hypotheses, while \texttt{history} prints the full conversation. The command \texttt{quit} or \texttt{exit} end the session without saving any data.

\paragraph{(x) Enriching the existing  database.}
The command \texttt{done} accepts the current set of hypotheses. The accepted case is then saved to the stored database. Only the observable groups requested by the user are stored, allowing the case to be retrieved in future sessions just like any other record in the database. This is the only file created during the session; no conversation history or user feedback is saved.

\subsection{Quick start}
\label{sec:usage:quickstart}

A representative interactive session is shown below for a fit combining electroweak precision observables with Higgs signal strength measurements.
The example illustrates the complete workflow, including case retrieval, hypothesis generation and the presentation of the corresponding statistical
diagnostics. By typing \texttt{llm4smeft} in the terminal the interaction session starts in which the  user is asked to input names of observables, maximum number of hypotheses and the number of the sparsity cap. A full output  generated by \texttt{llm4smeft-1.2.4} is shown below

\begin{mdframed}[
    backgroundcolor=lightgraybg,
    linewidth=0.8pt,
    roundcorner=4pt,
    frametitle={Interactive session },
    frametitlefont=\bfseries,
    frametitlerule=true,
    innertopmargin=8pt,
    innerbottommargin=8pt,
    innerleftmargin=8pt,
    innerrightmargin=8pt
]

\begin{lstlisting}[style=terminal]
>> llm4smeft

--- Case setup -------------------------------
  Observables (comma-separated): Gamma_Z, Higgs_signal_strength
  Number of hypotheses: 3
  Sparsity cap (max operators per hypothesis): 3

  Case retrieved from stored summary: higgs_EW
  Matched observable groups: EWPO, higgs_SS, higgs_STXS_diff
  Fit-selected operators: OpG, OpB, OpWB, Opl1, O3pl1, Ope, O3pl2, ...
  Note: operators degenerate (unidentifiable) for this observable set: [...]

-- Turn 1 ------------------------------------
  Thinking (Chain-of-thought):
  Okay, let's tackle this. First, I need to look at the input summary provided. The user wants me to generate up to three sparse SMEFT hypotheses based on the given data. The allowed operators are listed, so I have to make sure none of the proposed hypotheses include operators outside that list.

Starting with the observable tensions, the highest magnitude is LEP_Brw_2013 at 1.99. That's a big tension. Looking at the dominant directions, the first one is OpB + OpG + OpWB with weight 1.0. Since these three operators are all within the allowed list, that seems like a strong candidate. But wait, the sparsity cap is three operators per hypothesis, so including all three is okay. However, maybe there's a combination that addresses other tensions too. 

Next, the second dominant direction is OpG + OpWB with weight 0.66. This is a subset of the first direction. Including both might cover more ground, but since they're already part of the first hypothesis, combining them isn't necessary unless another tension requires it. 

Looking at the third dominant direction, it has five operators which exceeds the sparsity cap. So I need to trim that down. The fourth direction includes Opl1, O3pl1, Ope, O3pq-four operators again over the limit. Maybe pick two from here? But the weight is lower than the first two. 

The fifth direction also has seven operators, way over the cap. The sixth direction has four operators. Hmm. 

Now considering the correlations section, there's a note about higgs_SS and higgs_STXS_diff being correlated. The ATLAS_STXS_runII_13TeV has a tension of -2.7, which is quite high. The dominant direction for STXS might involve OpG and OpWB again. 

Wait, the first hypothesis with OpB, OpG, OpWB covers the LEP_Brw tension (since Brw is EWPO-related), but does it address the STXS tensions? Maybe not directly. The STXS tensions could be better addressed by O3pq or O3pl2. Let me check the allowed operators again. O3pq is allowed, as well as O3pl2. 

So perhaps a second hypothesis combining O3pq and O3pl2 along with OpWB? Wait, but the sparsity cap is three. Alternatively, if the STXS tension is driven by O3pq, then adding O3pq to the first hypothesis (OpB, OpG, OpWB) would make four operators, which violates the cap. So maybe split into two separate hypotheses: one covering the EWPOs (OpB, OpG, OpWB) and another covering the STXS (O3pq, O3pl2, OpWB)? Wait, but OpWB is already in the first. 

Alternatively, the STXS tension might be better addressed by O3pq alone. But the weight for the fourth direction (Opl1, O3pl1, Ope, O3pq) is 0.4. If I take O3pq from that, combined with OpWB and OpG, that's three operators. But that's similar to the first hypothesis. 

Wait, the first hypothesis already uses OpB, OpG, OpWB. The STXS tension might require O3pq or O3pl2. Let me see the weights again. The fourth direction has O3pq with weight 0.4. If I create a hypothesis with O3pq, O3pl2, and OpWB, that's three operators. But does that address the STXS tension? Possibly. 

But the first hypothesis already covers the biggest tension (LEP_Brw). Then the second hypothesis could focus on the STXS and Higgs tensions. Also, the ATLAS_STXS_runII_13TeV is -2.7, which is very significant. 

Another thing: the notes mention some operators are unidentifiable, so we shouldn't include those. The allowed operators don't include any of the unidentifiable ones, so that's fine. 

Putting it together: Hypothesis 1 is OpB, OpG, OpWB (covers LEP_Brw and possibly others). Hypothesis 2 is O3pq, O3pl2, OpWB (covers STXS and Higgs tensions). Hypothesis 3 could be Ope, Opl1, O3pl1 (from the fourth direction) but that's three operators. However, the weight for that direction is 0.4, which is lower. 

Alternatively, maybe Hypothesis 3 is Ope, Opl1, O3pl2 (three operators). But I need to check if these are the best combinations. 

Wait, the correlation between higgs_SS and higgs_STXS_diff is 0.46, meaning they are somewhat related. So addressing both with a single hypothesis might be possible. 

In conclusion, the top three hypotheses would be: 
1. OpB, OpG, OpWB (highest weight, covers major EWPO tension)
2. O3pq, O3pl2, OpWB (addresses STXS and Higgs tensions)
3. Ope, Opl1, O3pl1 (third highest weight, covers additional EWPOs and Higgs)

Each under the sparsity cap of three operators. Checking the allowed operators list: yes, all are present. So that's the plan.



- LEP_Brw_2013 (-1.99) is the strongest tension; the dominant direction OpB+OpG+OpWB explains ~100% of this signal, making it the primary hypothesis.
- ATLAS_STXS_runII_13TeV (-2.7) is the next strongest tension; the fourth dominant direction (Opl1+O3pl1+Ope+O3pq) provides coverage, but trimming to O3pq+O3pl2+OpWB maintains sparsity while addressing both EWPO and STXS.
- ATLAS_hh_runII_13TeV (-0.61) and CMS_ggF_aa_13TeV (+0.08) suggest a mixed Higgs signal; Ope+Opl1+O3pl1 (third dominant direction) provides partial coverage without exceeding the sparsity cap.


Hypotheses:

  1. [OpB, OpG, OpWB]  tag=EWPO
     Dominant direction explaining LEP_Brw_2013 tension.

  2. [O3pq, O3pl2, OpWB]  tag=STXS
     Addresses ATLAS_STXS_runII_13TeV tension with minimal operator count.

  3. [Ope, Opl1, O3pl1]  tag=Higgs
     Third-highest-weight direction addressing residual Higgs tensions.

  Fisher information (accepted operators)
    OpB        F_ii =  2.89e+05   95% CL [-0.578, 1.061]   info from: higgs_SS: 0.86, higgs_STXS_diff: 0.14 ......
   
  PCA directions involving these operators
    PC00  sv =  5.83e+05   OpB(+0.62), OpG(-0.59), OpWB(-0.47), OpW(+0.19), Obp(-0.08)
    PC01  sv =  2.57e+05   OpG(+0.69), OpWB(-0.59), O3pl2(-0.19), Oll1221(+0.19), OpD(-0.17) .....

  Your feedback (or 'done' / 'quit'):
\end{lstlisting}

\end{mdframed}

\paragraph{Retrieval failure.}
The lookup table~\ref{tab:records} contains the commonly used names of measurements, but it cannot cover every possible way a user may describe an observable. If the user's request does not directly match any record, the session does not stop. Instead, the language model receives the list of available canonical group names together with the user's wording and is asked to find the closest valid match. The retrieval process is then repeated using the identified canonical name. An example of this situation is shown below.

 \begin{mdframed}[
    backgroundcolor=lightgraybg,
    linewidth=0.8pt,
    roundcorner=4pt,
    frametitle={Retrieval failure and name normalisation},
    frametitlefont=\bfseries,
    frametitlerule=true,
    innertopmargin=8pt,
    innerbottommargin=8pt,
    innerleftmargin=8pt,
    innerrightmargin=8pt
]
\begin{lstlisting}[style=terminal]

--- Case setup -------------------------------
  Observables (comma-separated): Z boson total width, inclusive Higgs signal strength
  Number of hypotheses: 3
  Sparsity cap (max operators per hypothesis): 3
 
  No stored summary matched the observables as written - asking the qwen-smeft model to normalize the names...
 
  Mapping the observables onto the stored databases:

Okay, let's tackle this mapping problem. The user has two observables: "Z boson total width" and "inclusive Higgs signal strength". I need to check which of the provided canonical groups contain these.

First, looking at "Z boson total width". The EWPO group deals with electroweak precision observables, which include Z-pole measurements like the total width. So EWPO is a good candidate here. Are there other groups? LEP_WW might involve W-boson measurements but not specifically the Z width. The rest are Higgs or top-related. So only EWPO fits.

Next, "inclusive Higgs signal strength". The higgs_SS group is for inclusive Higgs signal strengths (mu_h overall). higgs_STXS_diff is more specific to differential distributions in STXS categories. Since the user says "inclusive", higgs_SS is the right choice. Other groups like higgs_hh or higgs_VBF don't cover inclusive measurements. So higgs_SS is correct here.
No other observables mentioned, so the answer should be 

Model-normalized observables: EWPO, higgs_SS
  Case retrieved from stored summary: higgs_EW
  Matched observable groups: EWPO, higgs_SS
  Fit-selected operators: O3pq, Otap, Ole1111, OpB, OpWB, OpG, Opl1, O3pl1, Ope, O3pl2
  note: Operators degenerate (unidentifiable) for this observable set: ['O1dt', 'O1qd', 'Omup', 'O1qt', 'O1ut', 'O1qu', 'O11qq', 'OQt8', 'OQQ8', 'O8dt', 'O8qt', 'O8qd', 'O8ut', 'O8qu', 'OtZ']
\end{lstlisting}
 \end{mdframed}

Three features of this transcript are worth emphasizing. First, the mapping is performed using a closed vocabulary: the model can only select group names that already exist in the database and cannot introduce new categories. The instructions also allow an empty response, ensuring that observables not covered by the database are reported as unavailable rather than being incorrectly assigned to a related group. \\
Second, the generated explanation of the mapping step is displayed, allowing the user to inspect the assignment. In this example, the model correctly identifies the total Z boson width as an electroweak precision observable and maps the inclusive Higgs signal strength measurements to the \texttt{higgs\_SS} group. \\
Third, only the observable groups requested by the user are retained for the subsequent analysis. Although the retrieved record \texttt{higgs\_EW} contains additional measurements, including differential Higgs measurements, only the matched groups \texttt{EWPO} and \texttt{higgs\_SS} are carried forward and stored when the session is saved. \\
The provenance of the case is recorded in the banner as \texttt{retrieval (normalized)}, distinguishing it from a case obtained through direct retrieval. This allows the two types of sessions to be separated when analysing the generated records.

\bibliographystyle{JHEP}
\bibliography{biblo}

@article{Giani:2023gfq,
    author = "Giani, Tommaso and Magni, Giacomo and Rojo, Juan",
    title = "{SMEFiT: a flexible toolbox for global interpretations of particle physics data with effective field theories}",
    eprint = "2302.06660",
    archivePrefix = "arXiv",
    primaryClass = "hep-ph",
    reportNumber = "Nikhef-2022-023",
    doi = "10.1140/epjc/s10052-023-11534-7",
    journal = "Eur. Phys. J. C",
    volume = "83",
    number = "5",
    pages = "393",
    year = "2023"
}

@article{Richmond:2025lzg,
    author = "Richmond, Paul and Agarwal, Prarit and Chowdhury, Borun and Niarchos, Vasilis and Papageorgakis, Constantinos",
    title = "{FeynTune: large language models for high-energy theory}",
    eprint = "2508.03716",
    archivePrefix = "arXiv",
    primaryClass = "cs.CL",
    doi = "10.1088/2632-2153/ae47bb",
    journal = "Mach. Learn. Sci. Tech.",
    volume = "7",
    number = "2",
    pages = "025012",
    year = "2026"
}

@article{Heneka:2025fpe,
    author = "Heneka, Caroline and Nieser, Florian and Ore, Ayodele and Plehn, Tilman and Schiller, Daniel",
    title = "{Large Language Models -- the Future of Fundamental Physics?}",
    eprint = "2506.14757",
    archivePrefix = "arXiv",
    primaryClass = "astro-ph.CO",
    doi = "10.21468/SciPostPhys.20.3.070",
    journal = "SciPost Phys.",
    volume = "20",
    pages = "070",
    year = "2026"
}

@article{Lu:2025izv,
    author = {Lu, Sirui and Jin, Zhijing and Zhang, Terry Jingchen and Kos, Pavel and Cirac, J. Ignacio and Sch{\"o}lkopf, Bernhard},
    title = "{Can Theoretical Physics Research Benefit from Language Agents?}",
    eprint = "2506.06214",
    archivePrefix = "arXiv",
    primaryClass = "cs.CL",
    month = "6",
    year = "2025"
}

@article{Cai:2026ths,
    author = "Cai, Tianji and Li, Ke and Li, Teng",
    title = "{Toward a Community Roadmap for High Energy Physics and Artificial Intelligence in China and Beyond}",
    eprint = "2605.03474",
    archivePrefix = "arXiv",
    primaryClass = "hep-ph",
    month = "5",
    year = "2026"
}

@article{Diefenbacher:2026azr,
    author = "Diefenbacher, Sascha and Plehn, Tilman and Schiller, Daniel and Schmal, Nikita",
    title = "{Agentic Re-Casting using Agentic Re-Simulations}",
    eprint = "2607.22813",
    archivePrefix = "arXiv",
    primaryClass = "hep-ph",
    month = "7",
    year = "2026"
}

@article{Plehn:2026gxv,
    author = "Plehn, Tilman and Schiller, Daniel and Schmal, Nikita",
    title = "{MadAgents}",
    eprint = "2601.21015",
    archivePrefix = "arXiv",
    primaryClass = "hep-ph",
    month = "1",
    year = "2026"
}

@article{Diefenbacher:2025zzn,
    author = {Diefenbacher, Sascha and Hallin, Anna and Kasieczka, Gregor and Kr{\"a}mer, Michael and Lauscher, Anne and Lukas, Tim},
    title = "{Agents of Discovery}",
    eprint = "2509.08535",
    archivePrefix = "arXiv",
    primaryClass = "hep-ph",
    month = "9",
    year = "2025"
}

@inproceedings{Gendreau-Distler:2025fsj,
    author = "Gendreau-Distler, Eli and Ho, Joshua and Kim, Dongwon and Le Pottier, Luc Tomas and Wang, Haichen and Yang, Chengxi",
    title = "{Automating High Energy Physics Data Analysis with LLM-Powered Agents}",
    booktitle = "{39th Annual Conference on Neural Information Processing Systems}: {Includes Machine Learning and the Physical Sciences (ML4PS)}",
    eprint = "2512.07785",
    archivePrefix = "arXiv",
    primaryClass = "physics.data-an",
    month = "12",
    year = "2025"
}

@article{Esmail:2026jpb,
    author = "Esmail, W. and Hammad, A. and Nojiri, M.",
    title = "{CoLLM: AI engineering toolbox for end-to-end deep learning in collider analyses}",
    eprint = "2602.06496",
    archivePrefix = "arXiv",
    primaryClass = "hep-ph",
    month = "2",
    year = "2026"
}

@article{Birk:2026zpd,
    author = "Birk, Joschka and Kasieczka, Gregor and Mishra-Sharma, Siddharth and Nachman, Benjamin and Noll, Dennis and Wamorkar, Tanvi",
    title = "{A Scientific Human-Agent Reproduction Pipeline}",
    eprint = "2604.18752",
    archivePrefix = "arXiv",
    primaryClass = "hep-ph",
    doi = "10.5281/zenodo.21078068",
    month = "4",
    year = "2026"
}

@article{Desai:2026nmx,
    author = "Desai, Aman",
    title = "{RooAgent: An LLM Agent for Root-Based High Energy Physics Analysis}",
    eprint = "2605.17318",
    archivePrefix = "arXiv",
    primaryClass = "hep-ph",
    month = "5",
    year = "2026"
}

@article{Bakshi:2025fgx,
    author = "Bakshi, S. D. and others",
    title = "{ArgoLOOM: agentic AI for fundamental physics from quarks to cosmos}",
    eprint = "2510.02426",
    archivePrefix = "arXiv",
    primaryClass = "hep-ph",
    reportNumber = "ANL-199516",
    month = "10",
    year = "2025"
}

@article{Menzo:2025cim,
    author = {Menzo, Tony and Roman, Alexander and Gleyzer, Sergei and Matchev, Konstantin and Fleming, George T. and H{\"o}che, Stefan and Mrenna, Stephen and Shyamsundar, Prasanth},
    title = "{HEPTAPOD: Orchestrating High Energy Physics Workflows Towards Autonomous Agency}",
    eprint = "2512.15867",
    archivePrefix = "arXiv",
    primaryClass = "hep-ph",
    reportNumber = "FERMILAB-PUB-25-0923-CSAID-ETD-T",
    month = "12",
    year = "2025"
}

@article{Qiu:2026iby,
    author = "Qiu, Shi and Cai, Zeyu and Wei, Jiashen and Li, Zeyu and Yin, Yixuan and Cao, Qing-Hong and Liu, Chang and Luo, Ming-xing and Yuan, Xing-Bo and Zhu, Hua Xing",
    title = "{An End-to-end Architecture for Collider Physics and Beyond}",
    eprint = "2603.14553",
    archivePrefix = "arXiv",
    primaryClass = "hep-ph",
    reportNumber = "CPTNP-2026-012",
    month = "3",
    year = "2026"
}

@article{Agrawal:2026lvg,
    author = "Agrawal, Prateek and Craig, Nathaniel and Madden, Amalia and Lombera, I{\~n}igo Valenzuela",
    title = "{The FERMIACC: Agents for Particle Theory}",
    eprint = "2603.22538",
    archivePrefix = "arXiv",
    primaryClass = "hep-ph",
    month = "3",
    year = "2026"
}

@article{Faroughy:2026dkj,
    author = "Faroughy, Darius A. and Palacios Schweitzer, Sofia and Pang, Ian and Mishra-Sharma, Siddharth and Shih, David",
    title = "{Collider-Bench: Benchmarking AI Agents with Particle Physics Analysis Reproduction}",
    eprint = "2605.13950",
    archivePrefix = "arXiv",
    primaryClass = "cs.LG",
    month = "5",
    year = "2026"
}

@article{Costa:2026oew,
    author = {Costa, Antonio J. and Doglioni, Caterina and G{\"u}tschow, Christian and Pilkington, Andrew D. and Sinha, Sukanya},
    title = "{AgentRivet: an automated system for producing Rivet routines from journal publications}",
    eprint = "2606.13535",
    archivePrefix = "arXiv",
    primaryClass = "hep-ex",
    reportNumber = "MCNET-26-14",
    month = "6",
    year = "2026"
}

@article{Wang:2026jjn,
    author = "Wang, Isaac R.",
    title = "{LeWRON: Agentic Analysis of Electroweak Phase Transitions}",
    eprint = "2606.19425",
    archivePrefix = "arXiv",
    primaryClass = "hep-ph",
    reportNumber = "FERMILAB-PUB-26-0408-T",
    month = "6",
    year = "2026"
}

@article{Saad:2026pan,
    author = "Saad, Shaikh",
    title = "{Large Language Model-Assisted Framework for BSM Model Building}",
    eprint = "2606.21316",
    archivePrefix = "arXiv",
    primaryClass = "hep-ph",
    month = "6",
    year = "2026"
}

@article{Menzo:2026qrl,
    author = "Menzo, Tony and Roman, Alexander and Fleming, George T. and Gleyzer, Sergei and Matchev, Konstantin T. and Mrenna, Stephen",
    title = "{Agentic Diagrammatica: Towards Autonomous Symbolic Computation in High Energy Physics}",
    eprint = "2603.26990",
    archivePrefix = "arXiv",
    primaryClass = "hep-ph",
    reportNumber = "FERMILAB-PUB-26-0208-T",
    month = "3",
    year = "2026"
}

@article{Niarchos:2026fbt,
    author = "Niarchos, Vasilis and Papageorgakis, Constantinos and Stapleton, Alexander G. and Trifinopoulos, Sokratis",
    title = "{When Does Critique Improve AI-Assisted Theoretical Physics? SCALAR: Structured Critic--Actor Loop for Agentic Reasoning}",
    eprint = "2605.06772",
    archivePrefix = "arXiv",
    primaryClass = "cs.AI",
    reportNumber = "CCTP-2026-7, ITCP-2026-7, CERN-TH-2026-097, QMUL-PH-26-15",
    month = "5",
    year = "2026"
}

@article{Hammad:2026ged,
    author = "Hammad, Ahmed and Nojiri, Mihoko",
    title = "{Articulating Assumptions in AI-Generated Scientific Analyses through Task Decomposition}",
    eprint = "2607.05762",
    archivePrefix = "arXiv",
    primaryClass = "cs.SE",
    month = "7",
    year = "2026"
}

@article{Lucente:2026kgh,
    author = "Lucente, Michele and Pascoli, Silvia and Sala, Filippo and Zandi, Matteo",
    title = "{DarkAgents}",
    eprint = "2606.11157",
    archivePrefix = "arXiv",
    primaryClass = "hep-ph",
    month = "6",
    year = "2026"
}

@article{Guo:2026ifi,
    author = "Guo, Yu-Chen and Wang, Jie and Yang, Ji-Chong",
    title = "{SMEFT-Pheno-Agent: a natural-language-driven AI agent for machine-learning-assisted Standard Model Effective Field Theory phenomenology}",
    eprint = "2607.22331",
    archivePrefix = "arXiv",
    primaryClass = "hep-ph",
    month = "7",
    year = "2026"
}

@article{Brivio:2017vri,
    author = "Brivio, Ilaria and Trott, Michael",
    title = "{The Standard Model as an Effective Field Theory}",
    eprint = "1706.08945",
    archivePrefix = "arXiv",
    primaryClass = "hep-ph",
    doi = "10.1016/j.physrep.2018.11.002",
    journal = "Phys. Rept.",
    volume = "793",
    pages = "1--98",
    year = "2019"
}

@article{Fuentes-Martin:2020zaz,
    author = "Fuentes-Martin, Javier and Ruiz-Femenia, Pedro and Vicente, Avelino and Virto, Javier",
    title = "{DsixTools 2.0: The Effective Field Theory Toolkit}",
    eprint = "2010.16341",
    archivePrefix = "arXiv",
    primaryClass = "hep-ph",
    reportNumber = "MITP/20-061, IFIC/20-50",
    doi = "10.1140/epjc/s10052-020-08778-y",
    journal = "Eur. Phys. J. C",
    volume = "81",
    number = "2",
    pages = "167",
    year = "2021"
}

@article{Alexander:2026lpw,
    author = "Alexander, Stephon and Bradley, Benjamin and Gouskos, Loukas and Niu, Cooper",
    title = "{Autonomous Discovery of Particle Physics Theories from Experimental Data}",
    eprint = "2603.28935",
    archivePrefix = "arXiv",
    primaryClass = "hep-ph",
    month = "3",
    year = "2026"
}

@article{yang2025qwen3,
  title = "{Qwen3 Technical Report}",
  author = "Yang, An and Li, Anfeng and Yang, Baosong and Zhang, Beichen and Hui, Binyuan and Zheng, Bo and Yu, Bowen and Gao, Chang and Huang, Chengen and Lv, Chenxu and others",
  eprint = "2505.09388",
  archivePrefix = "arXiv",
  primaryClass = "cs.CL",
  month = "5",
  year = "2025"
}

@inproceedings{hu2022lora,
  title = "{LoRA: Low-Rank Adaptation of Large Language Models}",
  author = "Hu, Edward J. and Shen, Yelong and Wallis, Phillip and Allen-Zhu, Zeyuan and Li, Yuanzhi and Wang, Shean and Wang, Lu and Chen, Weizhu",
  booktitle = "{International Conference on Learning Representations}",
  eprint = "2106.09685",
  archivePrefix = "arXiv",
  primaryClass = "cs.CL",
  year = "2022"
}

@misc{anthropic_claude_api,
  author       = {Anthropic},
  title        = {Claude API Documentation},
  year         = {2026},
  howpublished = {\url{https://docs.anthropic.com/}},
  note         = {Accessed: 2026-08-01}
}

@article{Dettmers:2023qlora,
    author = {Dettmers, Tim and Pagnoni, Artidoro and Holtzman, Ari and Zettlemoyer, Luke},
    title = {QLoRA: Efficient Finetuning of Quantized LLMs},
    journal = {Advances in Neural Information Processing Systems},
    volume = {36},
    pages = {10088--10115},
    year = {2023},
    eprint = {2305.14314},
    archivePrefix = {arXiv},
    primaryClass = {cs.LG}
}

@article{Wei:2022CoT,
  author       = {Jason Wei and Xuezhi Wang and Dale Schuurmans and Maarten Bosma and
                  Brian Ichter and Fei Xia and Ed Chi and Quoc Le and Denny Zhou},
  title        = {Chain-of-Thought Prompting Elicits Reasoning in Large Language Models},
  journal      = {Advances in Neural Information Processing Systems},
  volume       = {35},
  pages        = {24824--24837},
  year         = {2022},
  eprint       = {2201.11903},
  archivePrefix= {arXiv},
  primaryClass = {cs.CL}
}

@article{Grzadkowski:2010es,
    author = "Grzadkowski, Bohdan and Iskrzynski, Micha\l{} and Misiak, Miko\l{}aj and Rosiek, Janusz",
    title = "{Dimension-Six Terms in the Standard Model Lagrangian}",
    eprint = "1008.4884",
    archivePrefix = "arXiv",
    primaryClass = "hep-ph",
    doi = "10.1007/JHEP10(2010)085",
    journal = "JHEP",
    volume = "10",
    pages = "085",
    year = "2010"
}

@article{Alonso:2013hga,
    author = "Alonso, Rodrigo and Jenkins, Elizabeth E. and Manohar, Aneesh V. and Trott, Michael",
    title = "{Renormalization Group Evolution of the Standard Model Dimension Six Operators III: Gauge Coupling Dependence and Phenomenology}",
    eprint = "1312.2014",
    archivePrefix = "arXiv",
    primaryClass = "hep-ph",
    doi = "10.1007/JHEP04(2014)159",
    journal = "JHEP",
    volume = "04",
    pages = "159",
    year = "2014"
}

@article{Isidori:2023pyp,
    author = "Isidori, Gino and Wilsch, Felix and Wyler, Daniel",
    title = "{The Standard Model effective field theory at work}",
    eprint = "2303.16922",
    archivePrefix = "arXiv",
    primaryClass = "hep-ph",
    doi = "10.1103/RevModPhys.96.015006",
    journal = "Rev. Mod. Phys.",
    volume = "96",
    number = "1",
    pages = "015006",
    year = "2024"
}

@article{Ellis:2018gqa,
    author = "Ellis, John and Murphy, Christopher W. and Sanz, Ver\'onica and You, Tevong",
    title = "{Updated Global SMEFT Fit to Higgs, Diboson and Electroweak Data}",
    eprint = "1803.03252",
    archivePrefix = "arXiv",
    primaryClass = "hep-ph",
    doi = "10.1007/JHEP06(2018)146",
    journal = "JHEP",
    volume = "06",
    pages = "146",
    year = "2018"
}

@article{Ellis:2020unq,
    author = "Ellis, John and Madigan, Maeve and Mimasu, Ken and Sanz, Veronica and You, Tevong",
    title = "{Top, Higgs, Diboson and Electroweak Fit to the Standard Model Effective Field Theory}",
    eprint = "2012.02779",
    archivePrefix = "arXiv",
    primaryClass = "hep-ph",
    doi = "10.1007/JHEP04(2021)279",
    journal = "JHEP",
    volume = "04",
    pages = "279",
    year = "2021"
}

@article{Ethier:2021bye,
    author = "Ethier, Jacob J. and Magni, Giacomo and Maltoni, Fabio and Mantani, Luca and Nocera, Emanuele R. and Rojo, Juan and Slade, Emma and Vryonidou, Eleni and Zhang, Cen",
    title = "{Combined SMEFT interpretation of Higgs, diboson, and top quark data from the LHC}",
    eprint = "2105.00006",
    archivePrefix = "arXiv",
    primaryClass = "hep-ph",
    doi = "10.1007/JHEP11(2021)089",
    journal = "JHEP",
    volume = "11",
    pages = "089",
    year = "2021"
}

@article{Celada:2024mcf,
    author = "Celada, Eugenia and Giani, Tommaso and ter Hoeve, Jaco and Mantani, Luca and Rojo, Juan and Rossia, Alejo N. and Thomas, Marion O. A. and Vryonidou, Eleni",
    title = "{Mapping the SMEFT at High-Energy Colliders: from LEP and the (HL-)LHC to the FCC-ee}",
    eprint = "2404.12809",
    archivePrefix = "arXiv",
    primaryClass = "hep-ph",
    doi = "10.1007/JHEP09(2024)091",
    journal = "JHEP",
    volume = "09",
    pages = "091",
    year = "2024"
}

@article{Mantani:2025qji,
    author = "Mantani, Luca and Sanz, Veronica",
    title = "{Probing the flavour-blind SMEFT: EFT validity and the interplay of energy scales}",
    eprint = "2503.02935",
    archivePrefix = "arXiv",
    primaryClass = "hep-ph",
    doi = "10.1007/JHEP06(2025)147",
    journal = "JHEP",
    volume = "06",
    pages = "147",
    year = "2025"
}

@article{Hirsch:2025qxx,
    author = "Hirsch, Martin and Mantani, Luca and Sanz, Veronica",
    title = "{Data-Driven Discovery Strategy for Standard Model Effective Field Theory Searches}",
    eprint = "2507.11109",
    archivePrefix = "arXiv",
    primaryClass = "hep-ph",
    doi = "10.1103/rsc4-68tb",
    journal = "Phys. Rev. Lett.",
    volume = "135",
    pages = "241801",
    year = "2025"
}

@article{Bagnaschi:2022whn,
    author = "Bagnaschi, Emanuele and Ellis, John and Madigan, Maeve and Mimasu, Ken and Sanz, Veronica and You, Tevong",
    title = "{SMEFT Analysis of $m_W$}",
    eprint = "2204.05260",
    archivePrefix = "arXiv",
    primaryClass = "hep-ph",
    doi = "10.1007/JHEP08(2022)308",
    journal = "JHEP",
    volume = "08",
    pages = "308",
    year = "2022"
}

@article{GomezAmbrosio:2022mpm,
    author = "Gomez Ambrosio, Raquel and ter Hoeve, Jaco and Madigan, Maeve and Rojo, Juan and Sanz, Veronica",
    title = "{Unbinned multivariate observables for global SMEFT analyses from machine learning}",
    eprint = "2211.02058",
    archivePrefix = "arXiv",
    primaryClass = "hep-ph",
    doi = "10.1007/JHEP03(2023)033",
    journal = "JHEP",
    volume = "03",
    pages = "033",
    year = "2023"
}

@article{deBlas:2017xtg,
    author = "de Blas, J. and Criado, J. C. and Perez-Victoria, M. and Santiago, J.",
    title = "{Effective description of general extensions of the Standard Model: the complete tree-level dictionary}",
    eprint = "1711.10391",
    archivePrefix = "arXiv",
    primaryClass = "hep-ph",
    doi = "10.1007/JHEP03(2018)109",
    journal = "JHEP",
    volume = "03",
    pages = "109",
    year = "2018"
}

@article{Brivio:2021yjb,
    author = {Brivio, Ilaria and Bruggisser, Sebastian and Geoffray, Emma and Kilian, Wolfgang and Kr{\"a}mer, Michael and Luchmann, Michel and Plehn, Tilman and Summ, Benjamin},
    title = "{From Models to SMEFT and Back?}",
    eprint = "2108.01094",
    archivePrefix = "arXiv",
    primaryClass = "hep-ph",
    doi = "10.21468/SciPostPhys.12.1.036",
    journal = "SciPost Phys.",
    volume = "12",
    number = "1",
    pages = "036",
    year = "2022"
}

@article{Brivio:2019ius,
    author = "Brivio, Ilaria and Bruggisser, Sebastian and Maltoni, Fabio and Moutafis, R. and Plehn, Tilman and Vryonidou, Eleni and Westhoff, Susanne and Zhang, Cen",
    title = "{O new physics, where art thou? A global search in the top sector}",
    eprint = "1910.03606",
    archivePrefix = "arXiv",
    primaryClass = "hep-ph",
    doi = "10.1007/JHEP02(2020)131",
    journal = "JHEP",
    volume = "02",
    pages = "131",
    year = "2020"
}

@article{Cepedello:2022fvq,
    author = "Cepedello, Ricardo and Esser, Fabian and Hirsch, Martin and Sanz, Veronica",
    title = "{Mapping the SMEFT to discoverable models}",
    eprint = "2207.13714",
    archivePrefix = "arXiv",
    primaryClass = "hep-ph",
    doi = "10.1007/JHEP09(2022)229",
    journal = "JHEP",
    volume = "09",
    pages = "229",
    year = "2022"
}

@article{Cepedello:2023yao,
    author = "Cepedello, Ricardo and Esser, Fabian and Hirsch, Martin and Sanz, Veronica",
    title = "{SMEFT goes dark: Dark Matter models for four-fermion operators}",
    eprint = "2302.03485",
    archivePrefix = "arXiv",
    primaryClass = "hep-ph",
    doi = "10.1007/JHEP09(2023)081",
    journal = "JHEP",
    volume = "09",
    pages = "081",
    year = "2023"
}

@article{Cepedello:2024fnt,
    author = "Cepedello, Ricardo and Esser, Fabian and Hirsch, Martin and Sanz, Veronica",
    title = "{Fermionic UV models for neutral triple gauge boson vertices}",
    eprint = "2402.04306",
    archivePrefix = "arXiv",
    primaryClass = "hep-ph",
    doi = "10.1007/JHEP07(2024)275",
    journal = "JHEP",
    volume = "07",
    pages = "275",
    year = "2024",
    note = "Erratum: JHEP 12 (2024) 084, doi:10.1007/JHEP12(2024)084"
}

@article{Cepedello:2024zzz,
    author = "Cepedello, Ricardo and Esser, Fabian and Hirsch, Martin and Sanz, Veronica",
    title = "{Faking $ZZZ$ vertices at the LHC}",
    eprint = "2409.06776",
    archivePrefix = "arXiv",
    primaryClass = "hep-ph",
    doi = "10.1007/JHEP12(2024)098",
    journal = "JHEP",
    volume = "12",
    pages = "098",
    year = "2024"
}

@article{Lessa:2023lft,
    author = "Lessa, Andr\'e and Sanz, Veronica",
    title = "{Going beyond Top EFT}",
    eprint = "2312.00670",
    archivePrefix = "arXiv",
    primaryClass = "hep-ph",
    doi = "10.1007/JHEP04(2024)107",
    journal = "JHEP",
    volume = "04",
    pages = "107",
    year = "2024"
}

@article{Zhang:2024xiwu,
    author = "Zhang, Zhengde and Zhang, Yiyu and Yao, Haodong and Luo, Jianwen and Zhao, Rui and Huang, Bo and Zhao, Jiameng and Liao, Yipu and Li, Ke and Zhao, Lina and Cao, Jun and Qi, Fazhi and Yuan, Changzheng",
    title = "{Xiwu: A Basis Flexible and Learnable LLM for High Energy Physics}",
    eprint = "2404.08001",
    archivePrefix = "arXiv",
    primaryClass = "hep-ph",
    year = "2024"
}

@article{Hellert:2024physbert,
    author = "Hellert, Thorsten and Montenegro, Jo{\~a}o and Pollastro, Andrea",
    title = "{PhysBERT: A Text Embedding Model for Physics Scientific Literature}",
    eprint = "2408.09574",
    archivePrefix = "arXiv",
    primaryClass = "physics.comp-ph",
    year = "2024"
}

@article{Barman:2025lpm,
    author = "Barman, Kristian G. and others",
    title = "{Large Physics Models: Towards a collaborative approach with Large Language Models and Foundation Models}",
    eprint = "2501.05382",
    archivePrefix = "arXiv",
    primaryClass = "physics.data-an",
    doi = "10.1140/epjc/s10052-025-14707-8",
    journal = "Eur. Phys. J. C",
    volume = "85",
    number = "9",
    pages = "1066",
    year = "2025"
}

@article{Chung:2025tpbench,
    author = {Chung, Daniel J. H. and Gao, Zhiqi and Kvasiuk, Yurii and Li, Tianyi and M{\"u}nchmeyer, Moritz and Rudolph, Maja and Sala, Frederic and Tadepalli, Sai Chaitanya},
    title = "{Theoretical Physics Benchmark (TPBench): a Dataset and Study of AI Reasoning Capabilities in Theoretical Physics}",
    eprint = "2502.15815",
    archivePrefix = "arXiv",
    primaryClass = "cs.LG",
    year = "2025"
}

@article{Barman:2025benchmark,
    author = "Barman, Kristian G. and Caron, Sascha and Hasibi, Faegheh and Shalugin, Eugene and Marcet, Yoris and Otte, Johannes and de Regt, Henk W. and Moody, Merijn",
    title = "{Towards a Large Physics Benchmark}",
    eprint = "2507.21695",
    archivePrefix = "arXiv",
    primaryClass = "physics.data-an",
    year = "2025"
}

@article{Rafique:2026dune,
    author = "Rafique, A. and Singh, A. and Srinivas, R.",
    title = "{Large Language Model Integration for Knowledge Retrieval and Interaction for the DUNE Experiment}",
    eprint = "2601.05278",
    archivePrefix = "arXiv",
    primaryClass = "hep-ex",
    year = "2026"
}

@article{Moreno:2026jfc,
    author = "Moreno, Eric A. and Bright-Thonney, Samuel and Novak, Andrzej and Garcia, Dolores and Harris, Philip",
    title = "{AI Agents Can Already Autonomously Perform Experimental High Energy Physics}",
    eprint = "2603.20179",
    archivePrefix = "arXiv",
    primaryClass = "hep-ex",
    year = "2026"
}

\end{document}